\documentclass[11pt, article]{aastex}
\begin{document}

\shorttitle{The HiVIS Spectropolarimeter}
\shortauthors{Harrington et al.}

\title{The New HiVIS Spectropolarimeter and Spectropolarimetric Calibration of the AEOS Telescope}
\author{D. M. Harrington, J. R. Kuhn, and K. Whitman}
\affil{Institute for Astronomy, University of Hawaii, Honolulu-HI-96822}
\email{dmh@ifa.hawaii.edu}

\begin{abstract}

	We designed, built, and calibrated a new spectropolarimeter for the HiVIS spectrograph (R$\sim$ 12000-49000) on the AEOS telescope.  We also did a polarization calibration of the telescope and instrument.  We will introduce the design and use of the spectropolarimeter as well as a new data reduction package we have developed, then discuss the polarization calibration of the spectropolarimeter and the AEOS telescope.  We used observations of unpolarized standard stars at many pointings to measure the telescope induced polarization and compare it with a Zemax model.  The telescope induces polarization of 1-6\% with a strong variation with wavelength and pointing, consistent with the altitude and azimuth variation expected.  We then used scattered sunlight as a linearly polarized source to measure the telescopes spectropolarimetric response to linearly polarized light.  We then made an all-sky map of the telescope's polarization response to calibrate future spectropolarimetry.    

\end{abstract}


\section{Introduction}
	
	We will describe the high-resolution visible spectrograph (HiVIS) and the design of the spectropolarimeter.  We will then show some examples of new data reduction packages we have developed, and discuss the observations of the unpolarized and polarized sources.  We will combine these measurements into a polarization calibration model for the instrument and telescope and discuss the polarization calibration methods as they can be applied to future observations with the HiVIS spectropolarimeter.

\section{The AEOS Spectrograph}

\begin{figure}[!h,!t,!b]
\includegraphics[width=.9\linewidth]{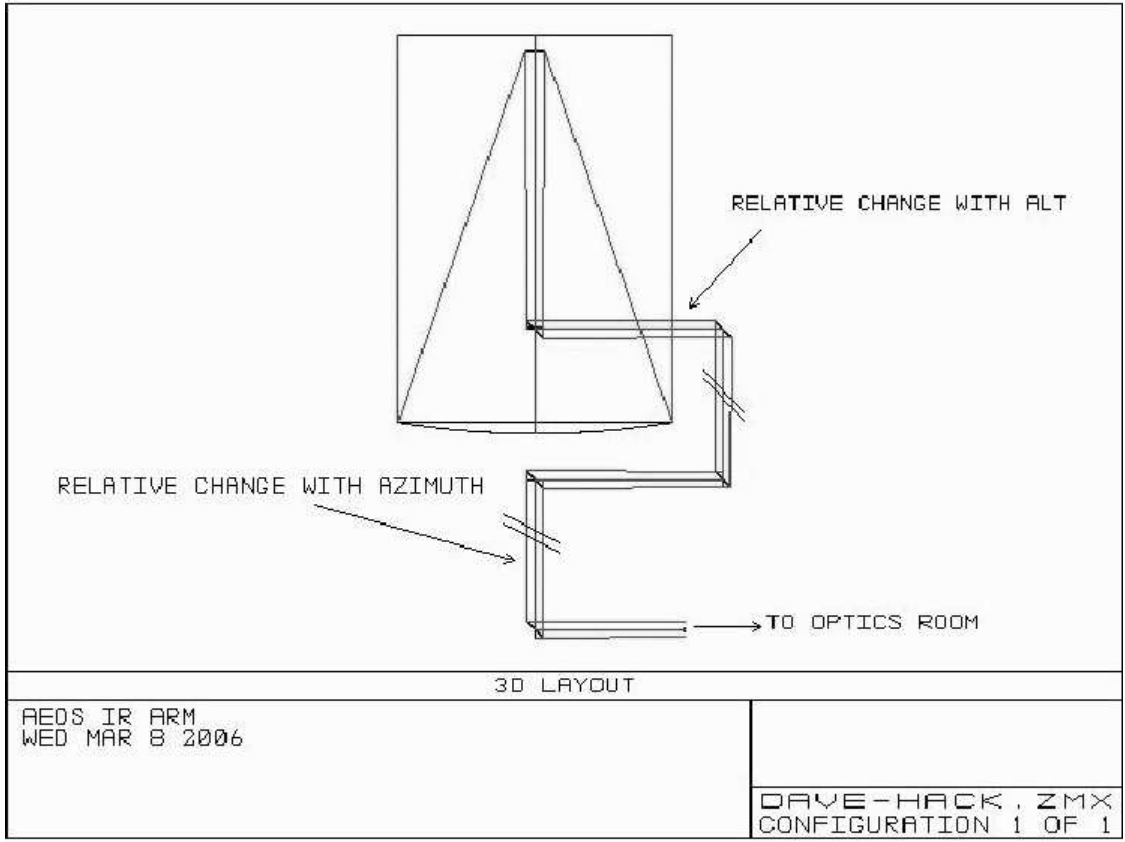}
\caption[aeostel]{\label{fig:aeostel}
A Zemax schematic of the coud\'e  telescope shown pointing at the zenith, shortened for illustration purposes.  These are the 7 coud\'e reflections (m1 to m7) before the optics room. }
\end{figure}

\begin{figure}[!h,!t,!b]
\includegraphics[width=\linewidth, height=\linewidth]{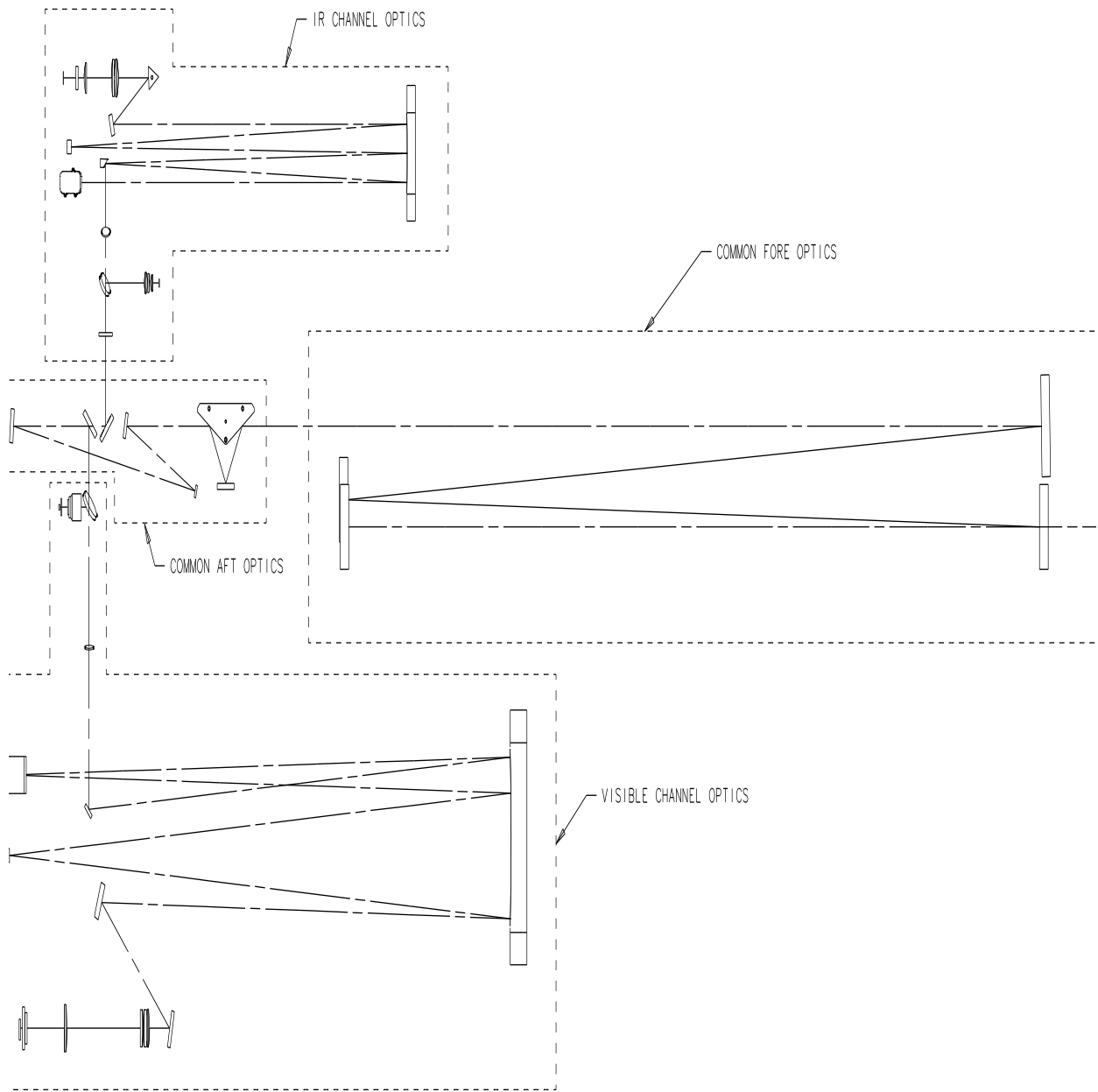}
\caption[Coude_layout]{\label{fig:Coude_layout}
This is a view of the AEOS optics room.  The common fore-optics are in the center of the figure, with the visible (bottom) and IR (top) arms of the spectrograph on either side.  The light enters the room from the center right after bouncing off the last coud\'e mirror, m7(not shown).}
\end{figure}

	The Advanced Electro-Optical System telescope (AEOS) is a 3.67m, altitude-azimuth telescope, located on Haleakala mountain, Maui, Hawaii.  The spectrograph uses the f/200 coud\'e  optical path, with seven reflections (five at $\sim45^\circ$ incidence) before coming into the optics room.  This is illustrated in figure \ref{fig:aeostel}, taken from the Zemax telescope design (industry standard optical design software).  The optics room, shown in figure \ref{fig:Coude_layout}, contains a visible (0.5-1.0$\mu$m) and IR (1.0-2.5$\mu$m) cross-dispersed echelle spectrographs.  The visible spectrograph has 2 gratings which cover 500-770nm and 640-1000nm at resolutions of 12800 to 48900.  The common fore optics includes three fold mirrors, collimator, K-cell (image rotator), fold-mirror, tip-tilt correction mirror, and a reimaging spherical mirror.  A movable pickoff mirror allows for switching between the visible and IR channels \citep{tho02, tho03}.
	
\begin{figure}[!h,!t,!b]
\includegraphics[ width=.9\linewidth]{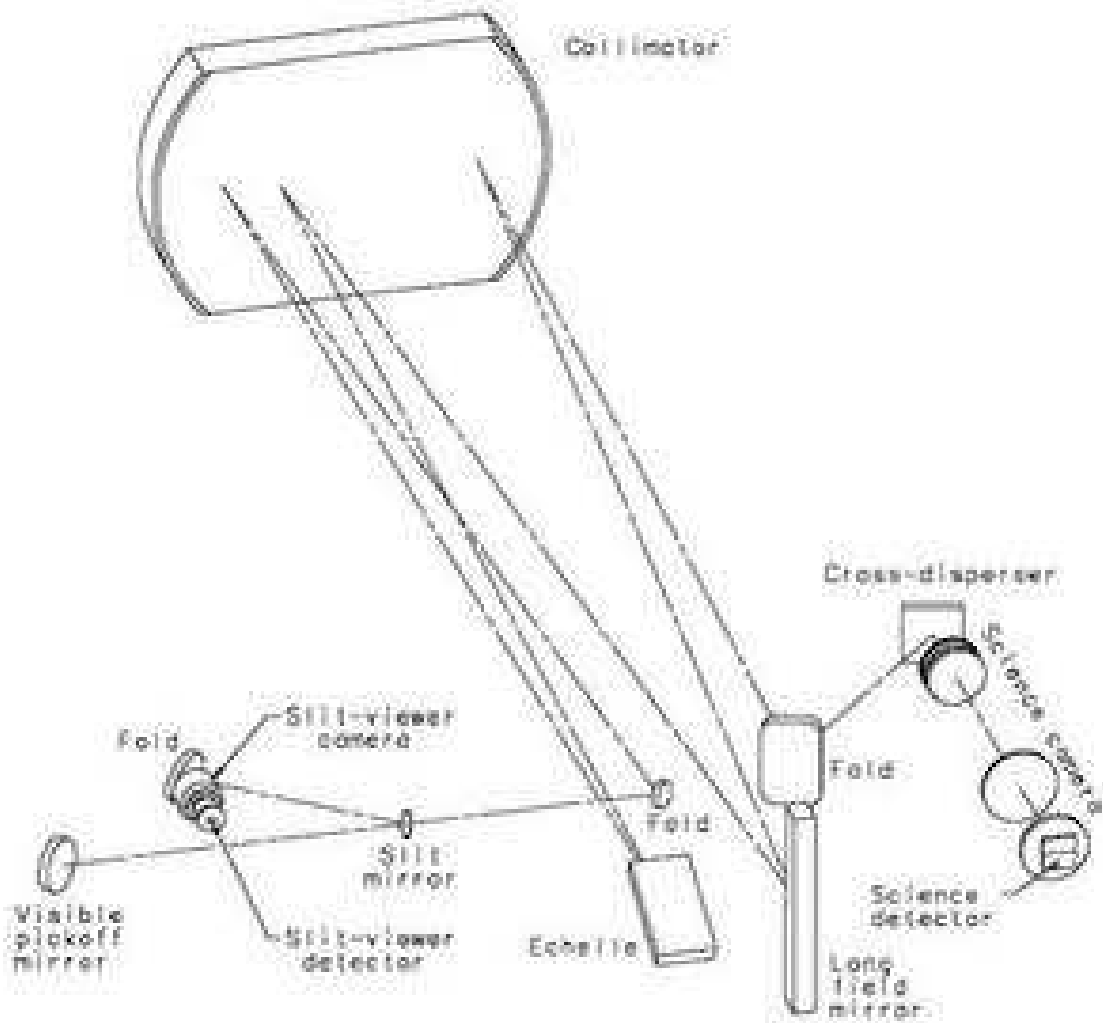}
\caption[Optical_bench.jpg]{\label{fig:opben}
This is a view of the AEOS visible spectrograph optical bench - The spectropolarimetry module is mounted just downstream of the slit, near the bottom left corner of the figure.}
\end{figure}

	The spectropolarimeter is designed for use with the visible arm of the AEOS spectrograph.  The visible arm is shown in figure \ref{fig:opben}.  From the pickoff mirror, shown in the bottom left corner of the figure, the beam passes through the slit in the slit-mirror, reflects off a fold mirror, the collimator, echelle, collimator, long mirror, collimator, fold mirror 2, and finally the cross disperser before being imaged on the CCD by a 5 lens camera.  There is a slit-viewer camera that is focused on the slit-mirror that allows real-time monitoring of the object.  The spectropolarimeter was mounted just downstream of the slit, near the image plane.  The plate scale is 0.16$\frac{``}{pixel}$. The cross disperser's 'Red' setting was used for all observations (nominally 6375-9680 $\AA$).  The spectrograph uses two 2K x 4K Lincoln Labs high resistivity CCD's mounted side by side with a 15 pixel gap to form a 4Kx4K array.  The CCD was set to rebin 2x2 so that the two output images are each 1024x2048 pixels.

\section{The Spectropolarimeter}

	The spectropolarimetry module for the AEOS spectrograph consists of a rotating achromatic half-wave plate and a calcite Savart Plate.  The Savart plate is two crossed and bonded calcite crystals which separate the incoming, uncollimated beam into two parallel, orthogonally polarized beams displaced laterally by 4.5mm.  The half-wave plate allows the rotation of the linear polarization of the incoming light to allow complete measurement of the linear polarization of the incident light.  The beams are displaced in the spatial direction (along the long axis of the slit).  

	In a Savart plate, the two beams (extraordinary and ordinary rays) are swapped between the two crossed crystals so both displaced rays have the same optical path length through the calcite.  This design contrasts with other spectropolarimeters which use Fresnel rhomb retarders and a Wollaston prism to separate the polarization states \citep{pet03, goo03, don04, don99, kaw99, don01, man03, goo95}.
	
\begin{figure}[htb]
\includegraphics[ width=.95\linewidth ]{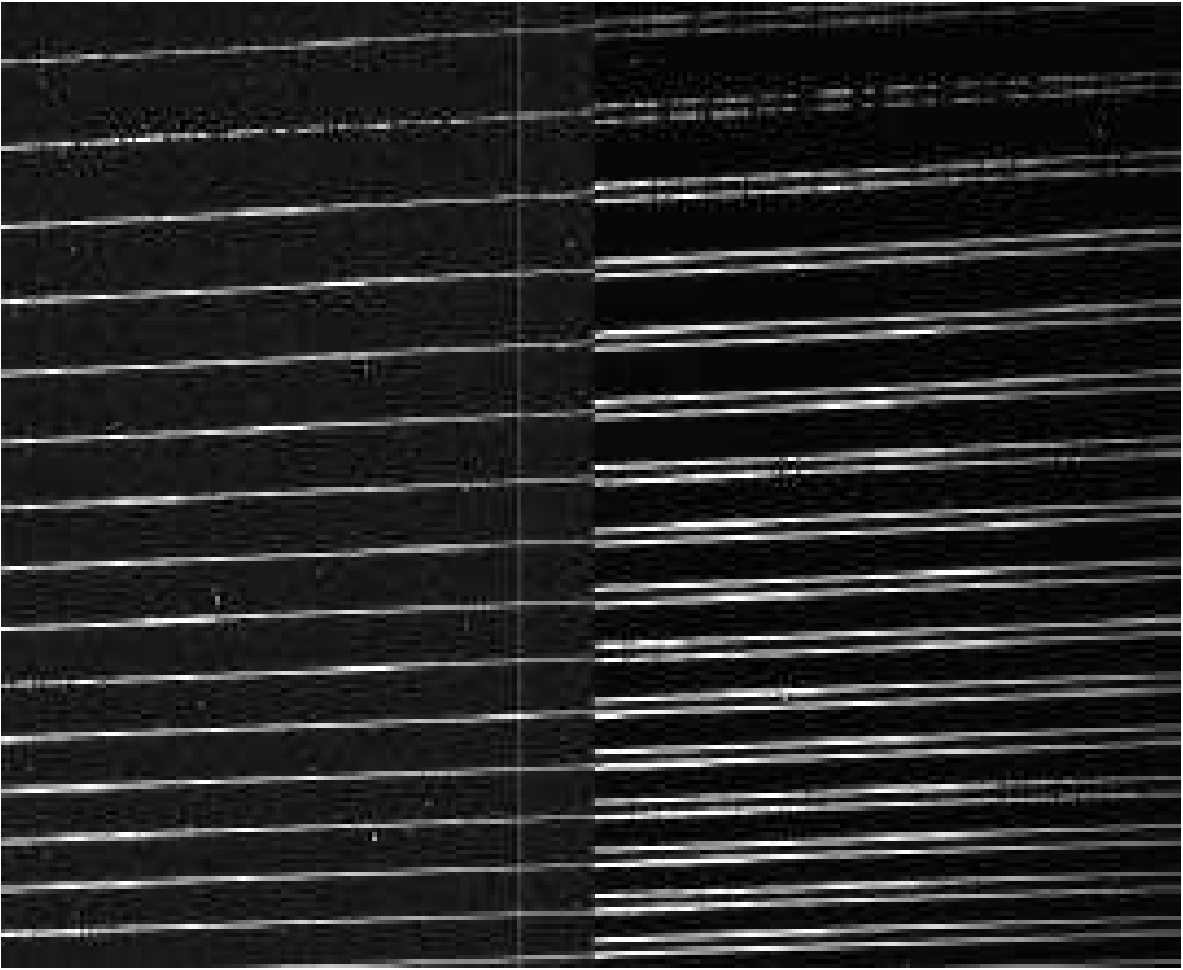}
\caption[chip]{\label{fig:chip}
The spectropolarimetry module splits each order into two orthogonally polarized orders.  The left image is a cropped raw image of a stellar spectra without the polarimeter (19 orders on the chip).  The right image is same region of the image with the same star with the polarimeter in place, doubling each order (2x19 orders).}
\end{figure}

	The small deviation in the image plane splits each spectral order detected on the CCD into two orthogonally polarized orders separated by a few pixels, from which orthogonally polarized spectra are extracted separately.  For example, the red cross disperser setting originally shows 19 orders across the CCD, but after insertion of the Savart plate, it shows 38 (2 polarizations for 19 orders) as illustrated in figure \ref{fig:chip}.

	  These components were added to the Zemax Inc. optical ray trace model of the spectrograph to assess the aberrations and determine the specifications of the optics in order to produce the proper polarized-order separation on the CCD ($\sim$25 pixels).  The aberrations induced by the waveplate and Savart plate were predicted to be smaller than half a pixel leading to insignificant image degradation .  A deviation of 4.5mm at the slit was found to separate the polarized-orders by $\sim$24.5 pixels on the CCD.  A 60mm Savart Plate produces this separation (0.075*60mm).  We then purchased and tested a Halbo Optics Inc. Savart plate with a 15mm aperture, a Bolder Vision Optik Inc. achromatic half-wave plate and a motorized, closed-loop rotation stage.  The half wave plate retardance was measured with a spectrograph and two crossed polarizers.  The retardance varied roughly quadratically with wavelength at the 5\% level from 500nm to 800nm (the useful range of our polarizers) and was 0.49 waves retardance at H$\alpha$.  The transmission of the Savart plate was above 95\% for these wavelengths. 

	After calibration and testing of the half-wave plate, Savart plate, and rotation stages, the instrument was assembled on the spectrograph on a movable plate which allows for the spectropolarimeter to be inserted or removed from the beam in under a minute.  We mounted a dekker upstream of the slit to reduce the slit length from 15'' to $\sim$7'' to minimize stray light introduced from the displacement.  This also assures polarized order separation on the CCD.   Flat fields taken after installation showed that the polarized orders were each $\sim$30 pixels across ($\sim$4") with a separation of 7 pixels between each polarized beam.  

	To test the equipment on the telescope, we mounted a linearly polarizing filter just upstream of the the spectropolarimeter, in order to feed 100\% linearly polarzed light to the spectropolarimeter, and detected 95 to 98\% polarization over the useful range of the polarizer (500-800nm, orders 0-12).  This shows we can detect linearly polarized light with the spectropolarimeter efficiently.

\section{Observing With AEOS - Measuring Linear Polarization}

	The Stokes parameters Q and U fully specify the linear polarization state of light \citep{col92}.  Q is the difference between horizontal and vertical polarizations.  U is the difference between +45$^\circ$ and -45$^\circ$ polarizations.  Typically, the fractional Stokes parameters q=Q/I  and u=U/I are presented to quantify the fraction of the light which is polarized.  These are measured by taking the fractional difference between the orthogonally polarized spectra. 

	The observing sequence with the AEOS spectrograph is to take spectra at 0, 22.5, 45, and 67.5 degrees rotation of the half-wave plate with respect to the axis of the Savart Plate, hereafter called orientation 1, 2, 3, and 4.  Incoming light linearly polarized at some angle to the fast axis of the half wave plate will have its linear polarization states rotated by twice that angle.  For example, incoming light polarized at 22.5$^\circ$ with respect to the wave plate axis will exit polarized at 45$^\circ$ (Stokes U $\rightarrow$ Q).  Since the axis of the Savart plate remains fixed, the rotation of the wave plate by 45$^\circ$ will move linearly polarized light from one polarized beam to the other, swapping their location on the CCD and reversing their sign in the fractional polarization calculation (Q $\rightarrow$ -Q).  This allows for cancellation of systematic errors (derivative, misalignment, CCD response, etc.) in the Stokes Q and U calculations.  The waveplate angle can be controlled to 50".
	
\begin{figure}[!h,!t,!b]
\includegraphics[ width=.95\linewidth ]{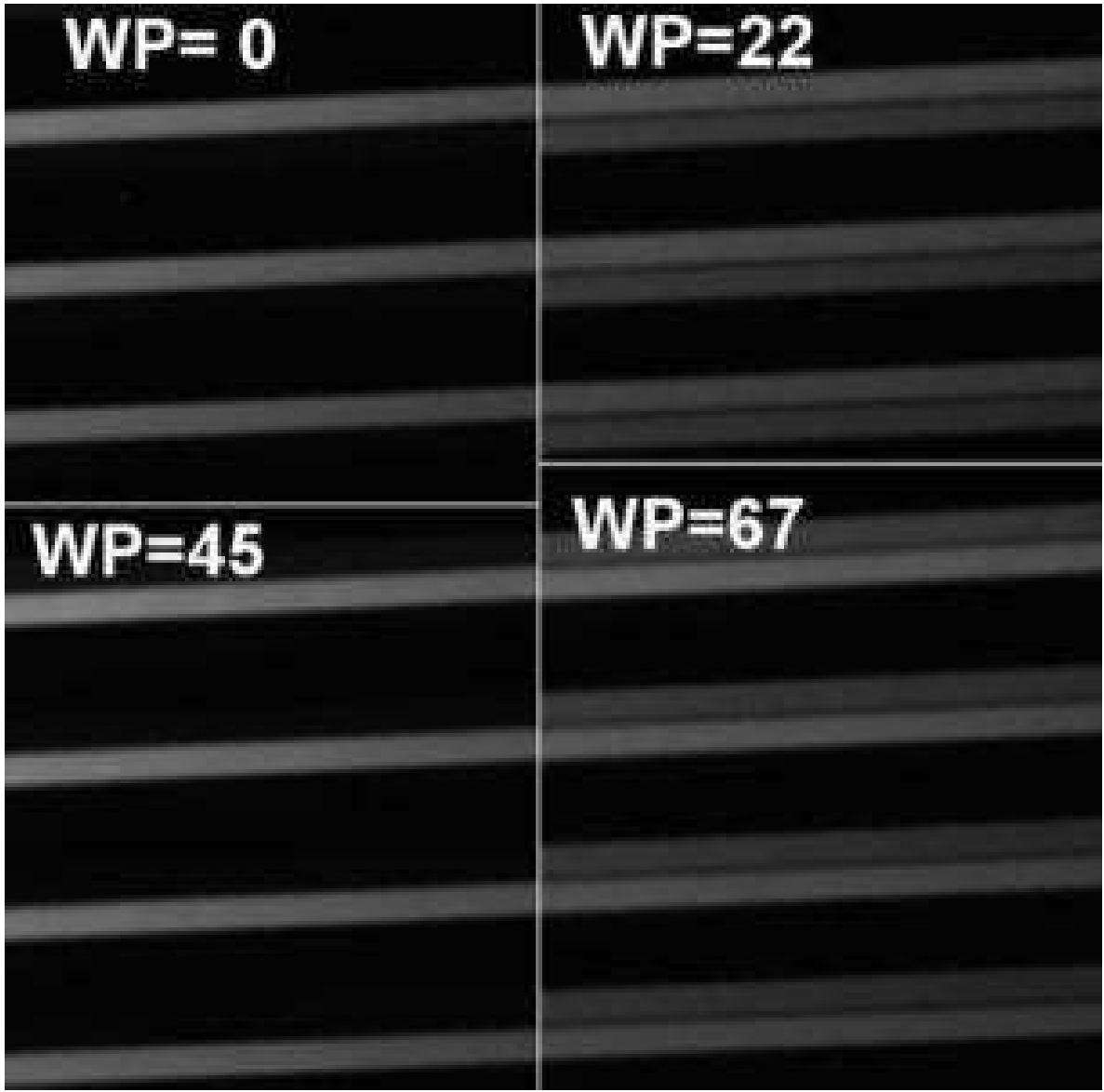}
\caption[4Ori]{\label{fig:4Ori}
A polarizer was mounted next to the slit to show how rotation of the waveplate swaps the order of polarized light.  Successive rotations of 22.5$^\circ$ take the light from the top order to the bottom order and back again.  Since a 45$^\circ$ rotation moved the light from the top to the bottom, instrumental systematic errors can be removed. Imperfect alignment with the Savart plate axis produces the slight asymmetry in the right hand panels.}
\end{figure}

	The observing sequence is illustrated in figure \ref{fig:4Ori}.  A linear polarizer was mounted upstream of the spectropolarimeter aligned with the waveplate's axis and a normal data set (4 waveplate orientations) was obtained.  Notice how the bright order is swapped (top to bottom) with a 45$^\circ$ rotation of the waveplate and how the 22$^\circ$ and 67$^\circ$ orientations show uniform intensity for both spectra.  The rotation of the waveplate will allow the linearly polarized light to be sampled by moving it from one polarized order to the other in a systematic way.

	The linear polarization measurements in this paper will be defined as follows:  Let a and b be the orthogonally polarized spectra (bottom and top respectively) with the waveplate at 0$^\circ$ and let c and d be the orthogonally polarized spectra with the waveplate at 45$^\circ$.  A measurement of a fractional Stokes q is defined as:

\begin{equation}
q=\frac{Q}{I}=\frac{1}{2}(\frac{a-b}{a+b} - \frac{c-d}{c+d})=\frac{1}{2}(q_{0^\circ}+q_{45^\circ})
\end{equation}

	Since a 45$^\circ$ rotation of the waveplate will swap the position of the polarized light incident along the original waveplate axis on the CCD (bottom to top or vice versa), but not the unpolarized light or linearly polarized light incident at 22.5$^\circ$ or 67.5$^\circ$, this measurement is only sensitive to Stokes q.  The subtraction of the two terms allows us to cancel systematic errors resulting from any misalignment of the polarized orders since each term is an independent measurement of Stokes q.  Measurement of Stokes U is the same formula, but with the waveplate at 22.5$^\circ$ and 67.5$^\circ$.  Since this is a fractional quantity, any correction applied equally to each order and to each chip, such as spectrophotometric calibrations, vignetting corrections or skyline subtractions, will not change the polarimetry in any significant way.  That is to say that no standard calibrations effect the spectropolarimetry.

\begin{figure}[!h,!t,!b]
\includegraphics[ width=.9\linewidth ]{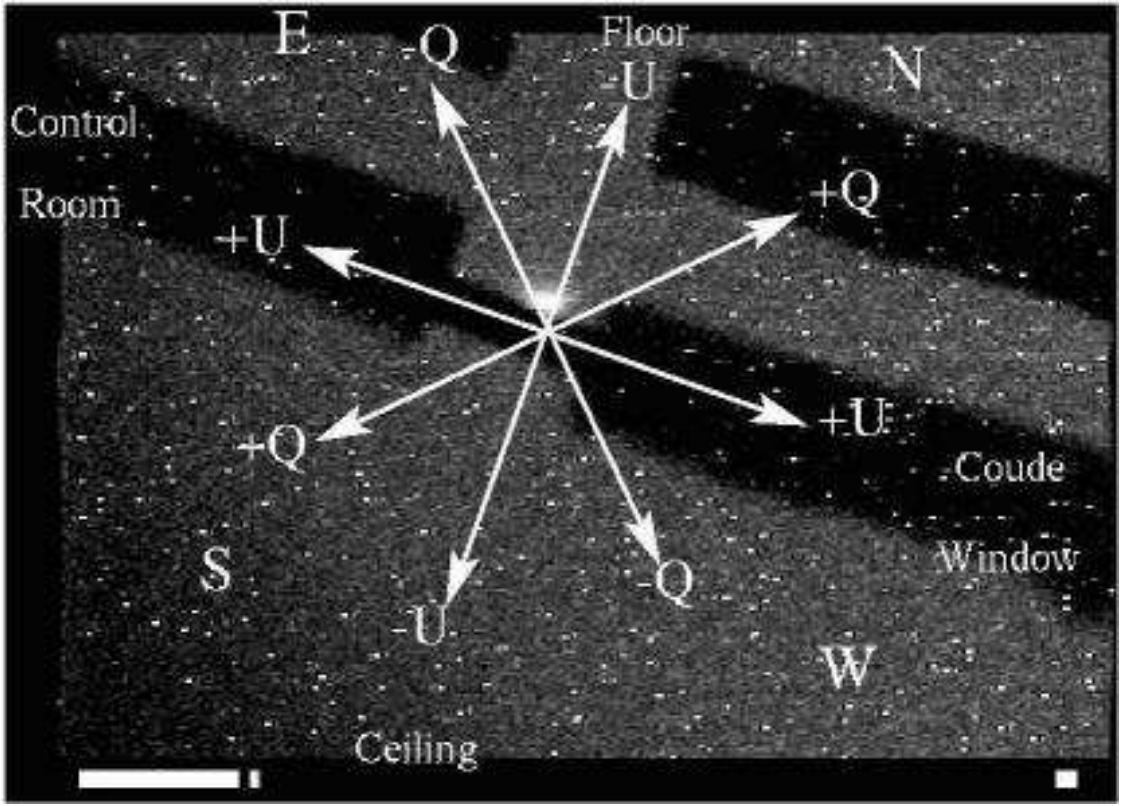}
\caption[SlitQU]{\label{fig:SlitQU}
The orientation of Q and U and the projection onto the sky as seen by the slitviewer camera at altitude 45$^\circ$ azimuth 225$^\circ$.  The Stokes parameters are fixed to the instrument, whereas the projection on the sky changes with time when the image rotator is not used.}
\end{figure}

	A frame from the slit-viewer camera is shown in figure \ref{fig:SlitQU} with the general geometry of the slit, dekker, and room as seen by the slit viewer camera.  The slit is parallel to the floor.  +U is parallel to the slit, and +Q is rotated $45^\circ$ counter clockwise, pointing to the upper right.  Since the image rotator was not used, the projection of the slit on the sky is a function of pointing and is not constant.

  	The simultaneous imaging of orthogonal polarization states also allows for a greater efficiency observing sequence and for reduction of atmospheric and systematic effects.   Since both polarized spectra are imaged simultaneously with identical seeing, a single polarization measurement (Q or U) can be done with a single image, and consistency between images is easy to quantify.  Using the fractional polarizations measured as the difference between the orthogonally polarized spectra divided by the sum (q and u) we can calculate the degree of polarization (P) and the position angle of polarization ($\theta$) as follows:

\begin{equation}
P = \sqrt{q^2 + u^2}
\end{equation}

\begin{equation}
\theta= \frac{1}{2}tan^{-1}\frac{q}{u}
\end{equation}

	The polarization angle is measured in the telescope frame and a projection of the slit's position angle onto the sky is done with the K-cell.  The AEOS image rotator was not used so that the absolute angle of the slit on the sky was not fixed.  However, the orientation of the K-cell mirrors in the optics room was fixed, making the polarization calibration much easier.

\begin{figure}[!h,!t,!b]
\includegraphics[ width=\linewidth , height=1.2\linewidth]{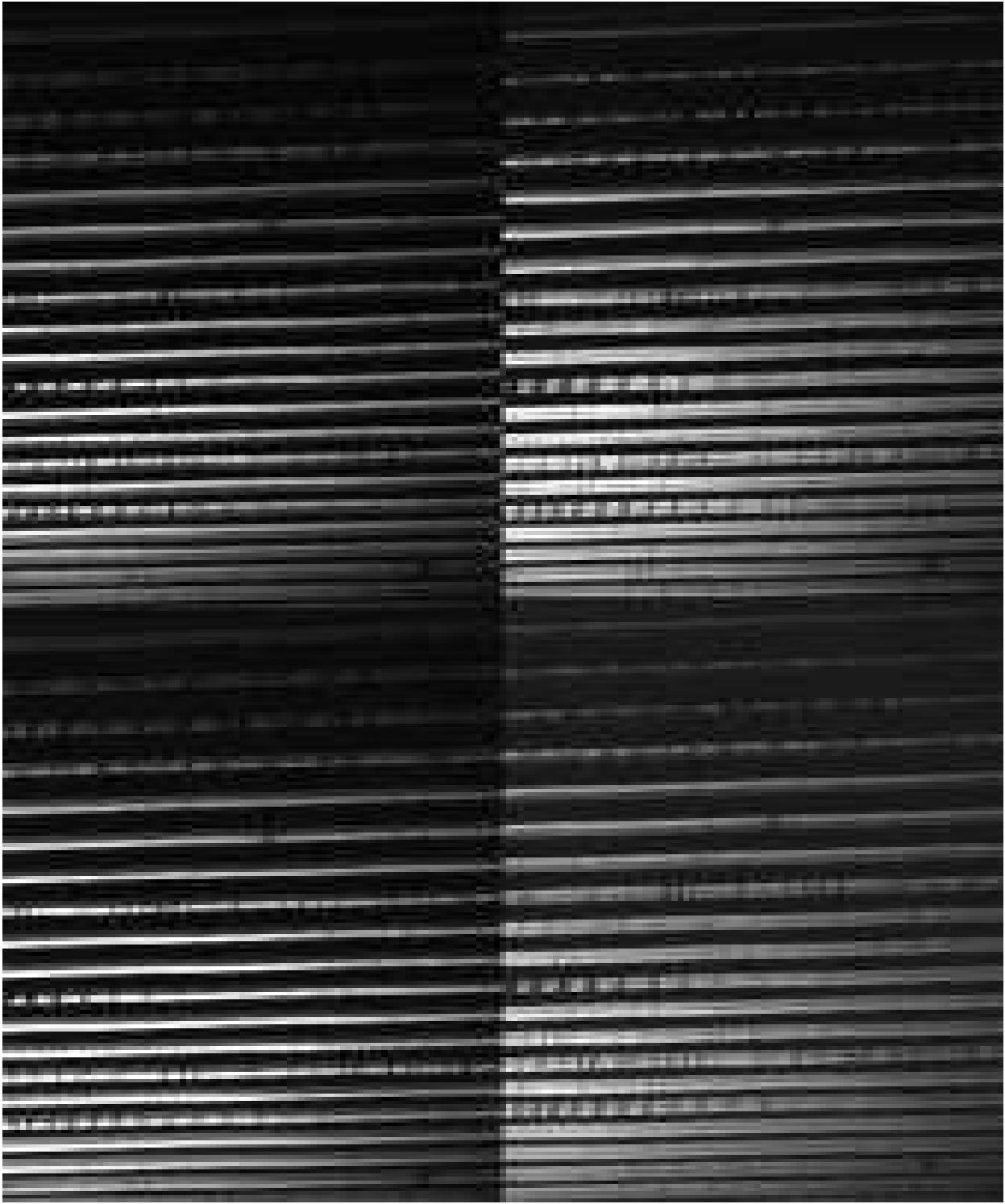}
\caption[S65_big]{\label{fig:skyset}
A scattered-sunlight data set at azimuth 180$\deg$ altitude 65$\deg$ taken at sunset.  Waveplate orientations are the same as figure \ref{fig:4Ori} with 0$^\circ$ in the upper left.}
\end{figure}

	We have written reduction scripts in IDL to reduce the data and compute the spectropolarimetry as described above.  Each spectral order is located by using a 2nd order polynomial fit to the intensity maximum of  a calibration star image.  This template for the order positions is then aligned with the data.  A median order profile is then obtained for each order.  A least-squares fit to this profile at each point along the order is used to calculate the intensity.  Standard flat fielding, bias subtraction, cosmic ray removal and wavelength calibration routines were also written.  We found that a 180 second exposure of an R=7.2 star gave spectra of S/N $\sim$50-200.  The S/N changes across each order, being highest where the disperser's efficiency is highest (eg. the middle left of each panel in figure \ref{fig:skyset}).  For example, the bottom order in figure \ref{fig:chip} goes from a S/N of 180 at 649nm in the bottom left to a S/N of 110 at 657nm in the bottom right, and the top order goes from a S/N of 110 at 956nm in the top left to a S/N of 60 at 968nm in the top right.    

	We will be neglecting circular polarization in this paper because the effects are typically much smaller than linear polarization, as will be discussed later, and we have no sensitivity to circularly polarized light.
	
	It should be noted that many spectropolarimeters show a polarimetric "ripple" effect caused by interference between multiple reflections in the waveplate (c.f. Aitken and Hough 2001).  We do not see such a ripple in our observations.  We have observed to 0.3\% polarimetric accuracy in a single frame at full resolution, and we are currently investigating higher S/N observations to verify the absence of this ripple.

\section{An Example: Spectropolarimetry of Scattered Sunlight}	

	Scattered sunlight is highly polarized at a scattering angle of 90$^\circ$.  We observed scattered twilight at many different pointings from June 26th to July 7th 2005 and a sample data set is shown in figure \ref{fig:skyset}.  Since the twilight filled the slit, the figure shows the width of each polarized order on the CCD ($\sim$27 pixels).  Wavelength increases to the right for each order, and towards the top with increasing order number.  The solar H$\alpha$ line is obvious in the lower right corner of each image.  The atmospheric A and B bands are on the lower left side in the 4th and 8th orders.

\begin{figure}[!h,!t,!b]
\includegraphics[ width=.8\linewidth, angle=90 ]{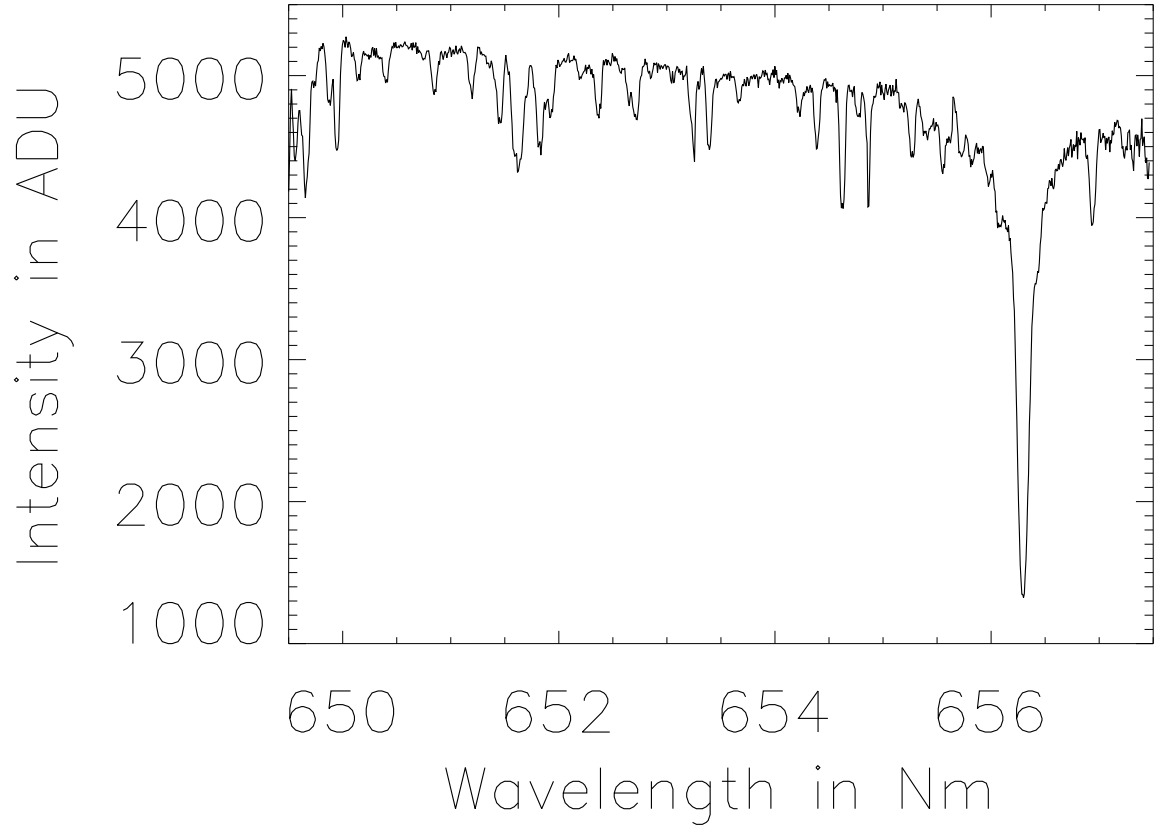}
\caption[solsp0]{\label{fig:solsp0}
The solar spectrum - spectra from the bottom polarization state for the first order (650nm - 657nm) as reduced with the new reduction software.}
\end{figure}

	As an illustration of the new data reduction software routines, a spectrum from one polarized order in a single twilight exposure is shown in figure \ref{fig:solsp0}.  The AEOS spectra match well with the spectra in a standard solar spectral atlas \citep{wal98}.  
	
\begin{figure}[!h,!t,!b]
\includegraphics[ width=.8\linewidth, angle=90]{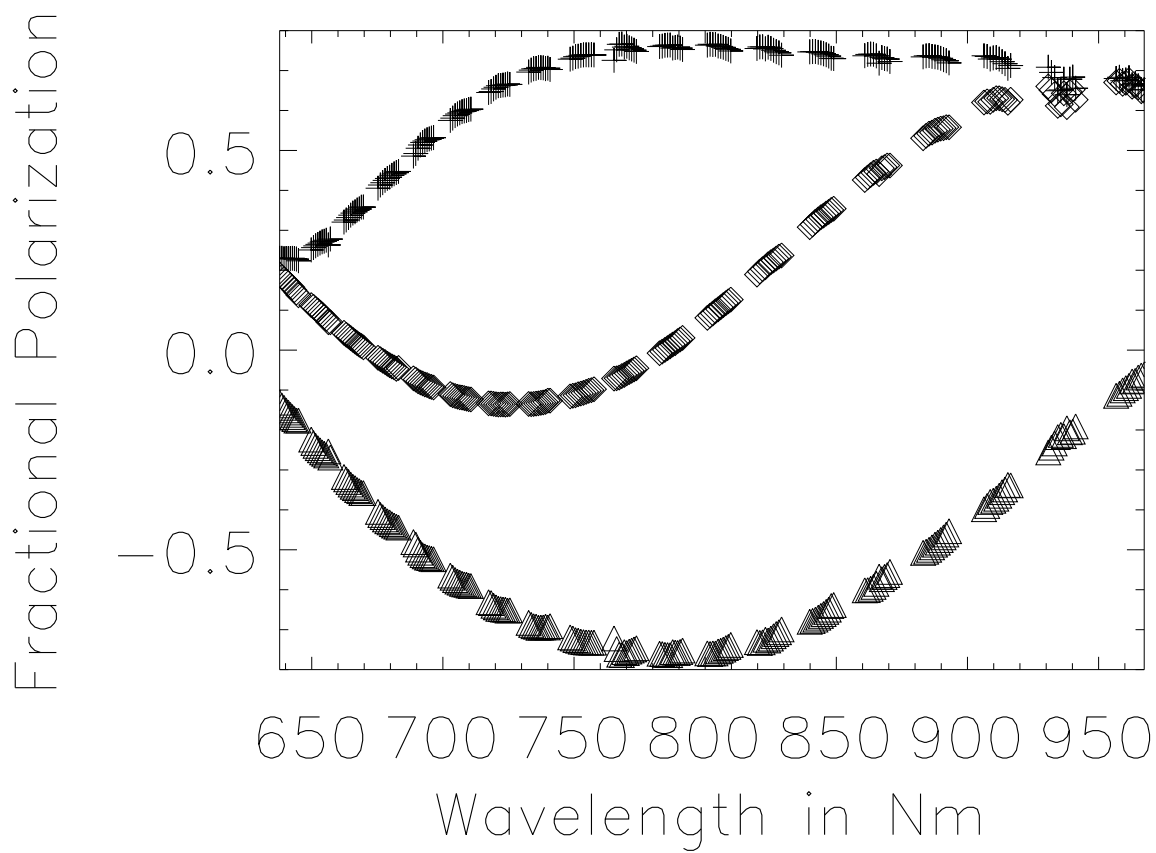}
\caption[sky65]{\label{fig:skys65}
Scattered sunlight polarization at azimuth of 180 and elevation 65.  Polarization of scattered sunlight is maximum at 90$^\circ$ scattering angle and at sunset, this is the arc from North to South through the Zenith.  The polarization spectra have been averaged 200:1 for ease of plotting.  The symbols are: q=$\Diamond$,  u=$\triangle$,  P=$\sqrt{Q^2+U^2}$=+.  P is the top curve which begins at 25\% at 650nm, rises to 70\% by 750nm and stays flat until 950nm.  Stokes q and u are the more sinuosoidal with a strong rotation of the plane of polarization ($\frac{1}{2}tan^{-1}\frac{q}{u}$).}
\end{figure}

		The polarization reduction package was applied to a twilight data set and the results are shown in figure \ref{fig:skys65}.  The polarization spectra have been averaged to 200 times lower spectral resolution for ease of plotting. The top curve is the degree of polarization which starts at around 25\% at 650nm and rises to 70\% by 700nm  and stays flat until 950nm.  Another interesting feature is the rotation of the plane of polarization across the CCD.   Notice how stokes q and u track each other.  At 650nm the magnitude of both q and u is 10\%.  At 800nm, q=0 and u=70\%.  Since the angle of polarization is $\frac{1}{2}tan^{-1}\frac{q}{u}$, this angle has rotated by  $90^\circ$.  The scattered sunlight observations will be discussed in more detail below.

\section{Polarization Calibration - The AEOS Model}

	The next step in the instrument development was calibrating the instrumental polarization response.  Absolute polarization at all wavelengths requires very careful calibration of the telescope.  Internal optics can induce or reduce polarization in the beam and also rotate the plane of polarization (Q $\rightarrow$ U).  All of these effects are functions of wavelength, altitude and azimuth.  Since AEOS is an  alt-az telescope, the mirror positions change as the telescope tracks, inducing polarization effects with pointing.  For AEOS, there are five $\sim45^\circ$ reflections before the Coud\'e  window, two of which change their relative orientation, shown in figure \ref{fig:aeostel}.  The first rotation is because the tertiary mirror is tied to the altitude axis while the next mirror, m4, sits on the mount.  The second relative rotation is because the mirror that sends the light to the coud\'e  mirror, m5, is tied to the azimuth axis while the coud\'e  room mirror, m6, is fixed.  
	
	Absolute spectropolarimetric calibration is difficult due to a lack of spectropolarimetrc standards.  Usually one observes polarimetric standards that have a precise value averaged over some bandpass \citep{ber82, hsu82, bas88, cla94, gil03}.  The calibration and creation of a telescope model is done by measuring unpolarized standard stars and polarized sources at many pointings to calibrate the effects of the moving mirrors.  
	
	If we ignore circular polarization, a simple and self consistent way of describing the telescopes response is a 3 variable system - 1) the difference in absorption between 2 orthogonal directions (mirror reflection axes), 2) the angle of those directions with respect to some fixed axis and 3) the wavelength dependence of those two parameters, all of which are functions of altitude and azimuth.  This is analogous to describing the telescopes response as an elliptical polarization at each pointing and each wavelength \citep{col92}.  In general, the telescope will have different responses to linearly polarized light since the absorption coefficients of the optical components vary with wavelength and the differential absorption is a function of the angle of the incident polarization with respect to the mirror/component axes.  While our models can calculate the full Mueller matrix as functions of altitude, azimuth, and wavelength, we cannot measure all of these terms.  We can, however, measure certain parts of the Mueller matrix given proper sources (polarized standards, unpolarized standards, and scattered sunlight).  

	Unpolarized standard stars are the first tool we used to calibrate the telescope.  Observations of these standards at as many pointings as possible allows one to construct a model of the telescopes response to unpolarized light and to quantify the telescope-induced polarization (but not the depolarization or plane rotation which requires a linearly polarized standard).

	Polarized sources were the last calibration source we used, and the most difficult.  Using a polarized source allows one to measure the depolarization and rotation of the plane of polarization, all of which are, in principle, orientation dependent.  Polarized standard stars are polarized in their continuum light, but the constant value quoted in standard references is an average polarization in a relatively line-free, wide bandpass image.  The degree of polarization is usually not more than a few percent making instrument calibration difficult with these sources.  A calibration method which we will explore below, is to use twilight (scattered sunlight) as a linearly polarized source with a roughly known degree of polarization.  

\subsection{Flat Field Polarization: Fixed Optics Polarization}

\begin{figure}[!h,!t,!b]
\includegraphics[ width=.8\linewidth, angle=90]{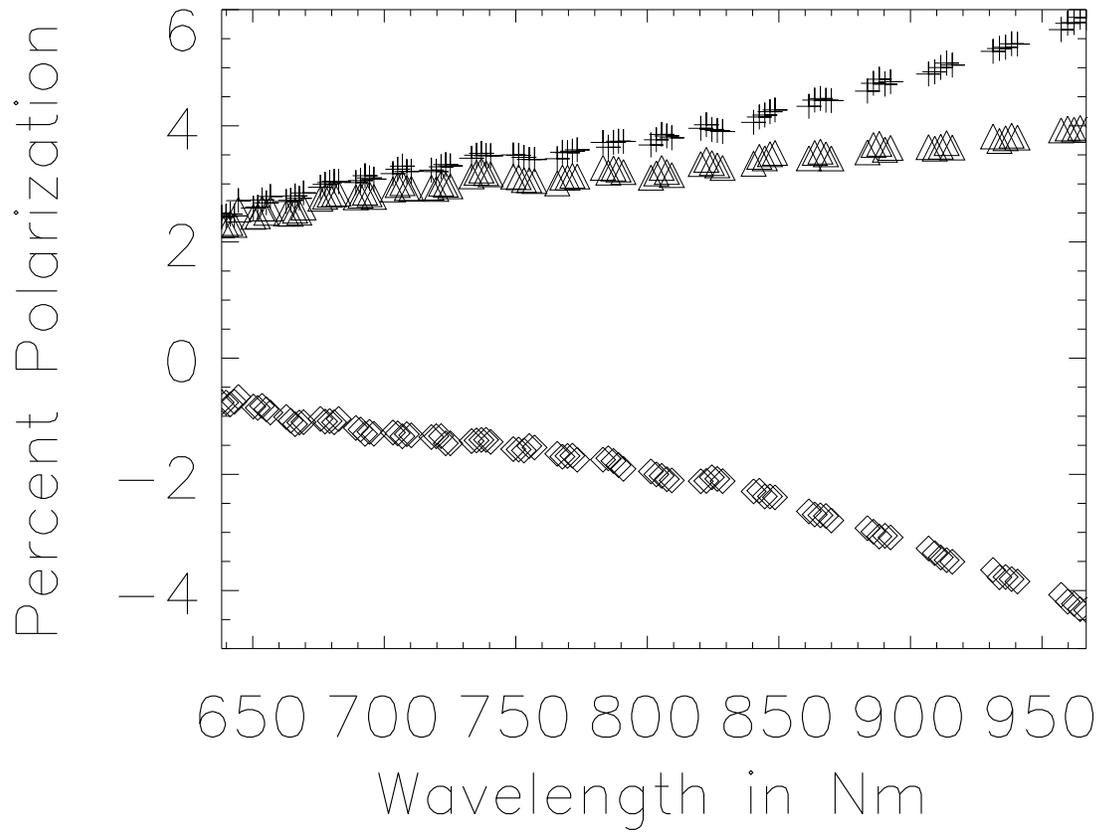}
\caption[flatpol]{\label{fig:flatpol}
Flat field polarization analysis, averaged 200:1 for clarity. Stokes q is shown as $\diamond$ , u is shown as $\triangle$, and P is shown as +.  The polarization (top curve) begins at 650nm being dominated by u (middle curve), but begins to rise as -q (bottom curve) gets larger.}
\end{figure}

	All of the moving mirrors in the AEOS telescope are located upstream of the polarizing optics.  As an illustration of the polarization induced by the reflections between the halogen flat field lamp ($T_{eff} \sim 2900K$) and the spectropolarimeter, figure \ref{fig:flatpol} shows the polarization reduction applied to a set of flat field frames with the 4 waveplate orientations.   The flat field lamp is located just after the Coud\'e  window, on the far right in figure \ref{fig:Coude_layout}.

	The flat fields show a significant polarization which changes with wavelength.  At 650nm, the induced polarization is dominated by the +U term, but the U and -Q terms become equal at 900nm. The position angle varies almost linearly across all wavelengths while the degree of polarization rises from 2\% at 650nm to 6\% at 950nm.  This illustrates the polarization induced by the 11 fixed reflections in the optics room.

\subsection{Unpolarized Standard Stars - Telescope Mirror Induced Polarization}

\begin{figure}[!h,!t,!b]
\includegraphics[ width=.8\linewidth, angle=90]{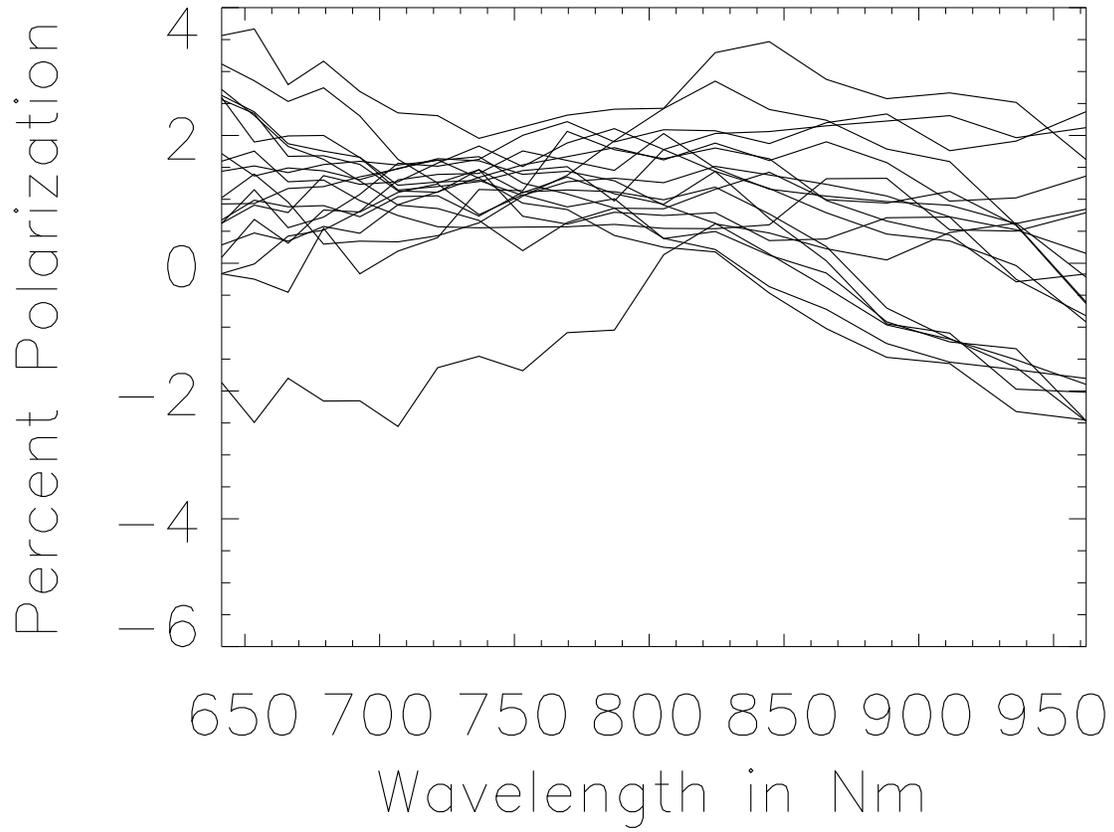}
\caption[all_unpol_q]{\label{fig:all_unpol_q}
Stokes q, rebinned 1000:1, for the unpolarized standard star observations in table \ref{upobs}.  Most observations show similar trends with wavelength (q falling).}
\end{figure}

\begin{figure}[!h,!t,!b]
\includegraphics[ width=.8\linewidth, angle=90]{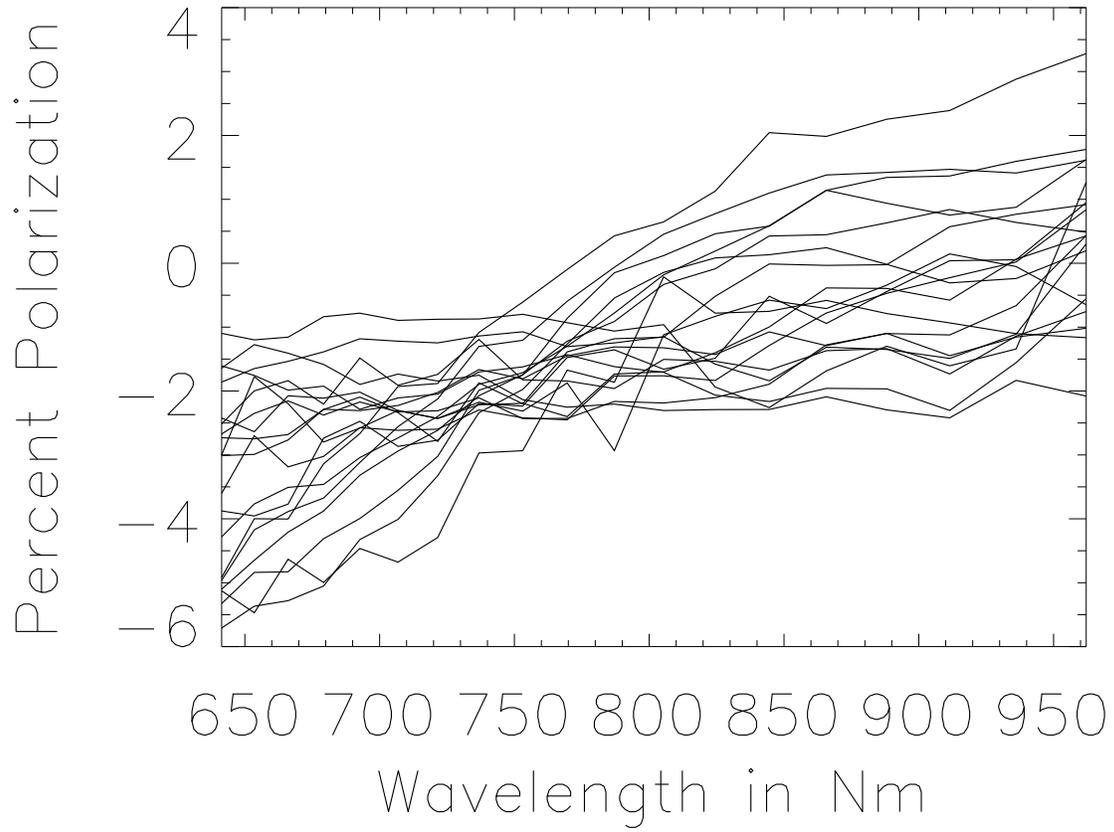}
\caption[all_unpol_u]{\label{fig:all_unpol_u}
Stokes u, rebinned 1000:1, for the unpolarized standard star observations in table \ref{upobs}.  They all show similar trends with wavelength (u rising).}
\end{figure}

	Many unpolarized standards have been observed in November of 2004 and in July of 2005 totaling 19 data sets (152 spectra).   A plot of q and u for all the unpolarized standard stars, where each 1000-pixel order has been rebinned to 1 data point,  is shown in figures \ref{fig:all_unpol_q} and \ref{fig:all_unpol_u}. They show the induced polarization as a function of pointing and wavelength.  There is a fairly consistent trend between all stars where q goes from 0-3\% at 650nm falling to between -2\% to +2\% at 950nm, and u starts between 0\% and -6\% at 650nm and then rises to between -2\% to +3\%.

   Calibration of the induced polarization to any future observations is done by observing the unpolarized standard stars at different pointings and creating a map of the telescopes response by projecting these measurements in alt-az space.  This map can then be interpolated to any pointing one wishes to calibrate.  The ultimate aim being to have a high resolution map of the telescope's polarization response.

\subsection{Examples of Unpolarized Standard Star Spectropolarimetry - Changes With Pointing}

\begin{figure}[!h,!t,!b]
\includegraphics[ width=.8\linewidth, angle=90]{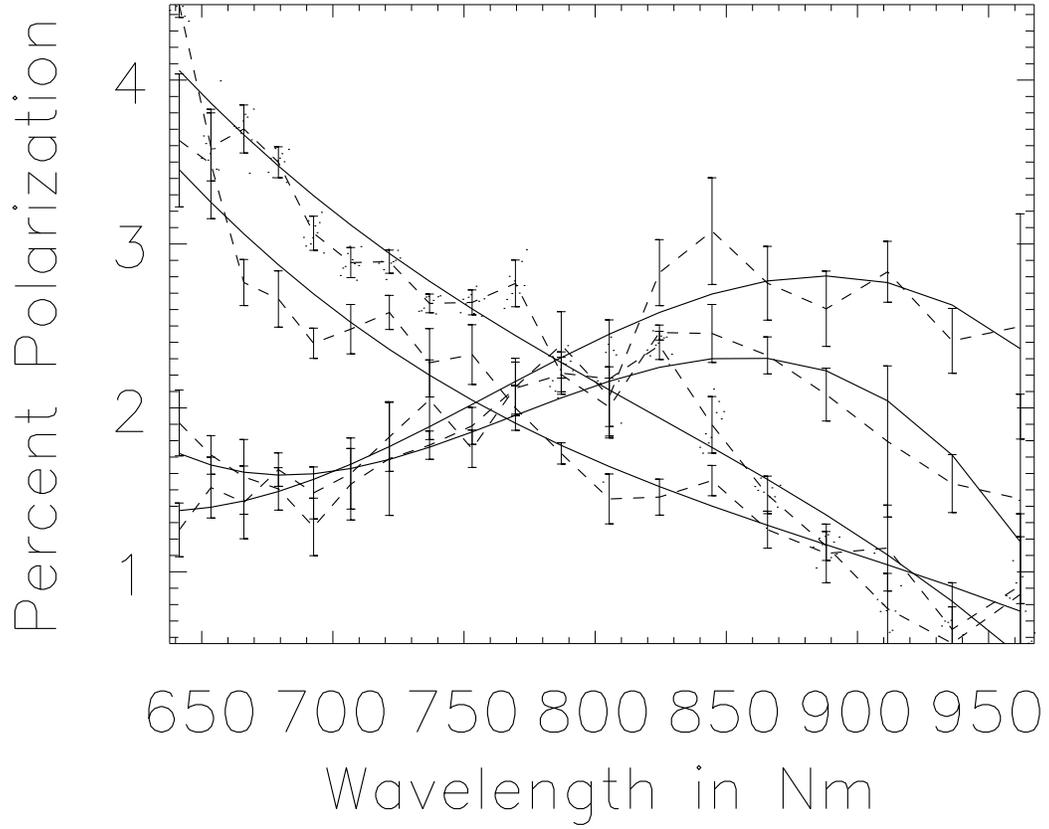}
\caption[pc]{\label{fig:125pc}
The degree polarization in percent for the unpolarized standard star HD125184. The progression in time at 650nm from the bottom curve to the top curve is 7:20, 8:00, 5:55, 5:45UT.}
\end{figure}

\begin{figure}[!h,!t,!b]
\includegraphics[ width=.8\linewidth, angle=90]{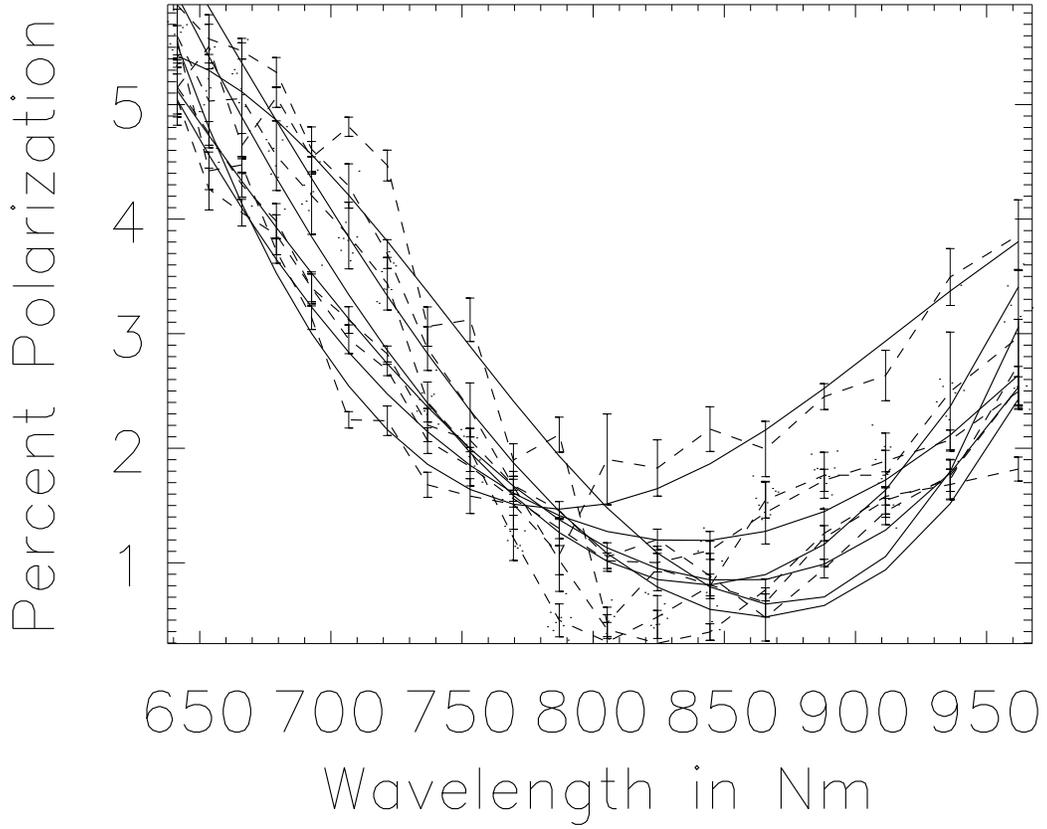}
\caption[pc]{\label{fig:114pc}
The degree of polarization in percent for the unpolarized standard star HD114710.  The progression in time is not very significant.  The curve that comes above the rest from 800nm to 950nm is at time 7:00UT.}
\end{figure}

\begin{figure}[!h,!t,!b]
\includegraphics[ width=.8\linewidth, angle=90]{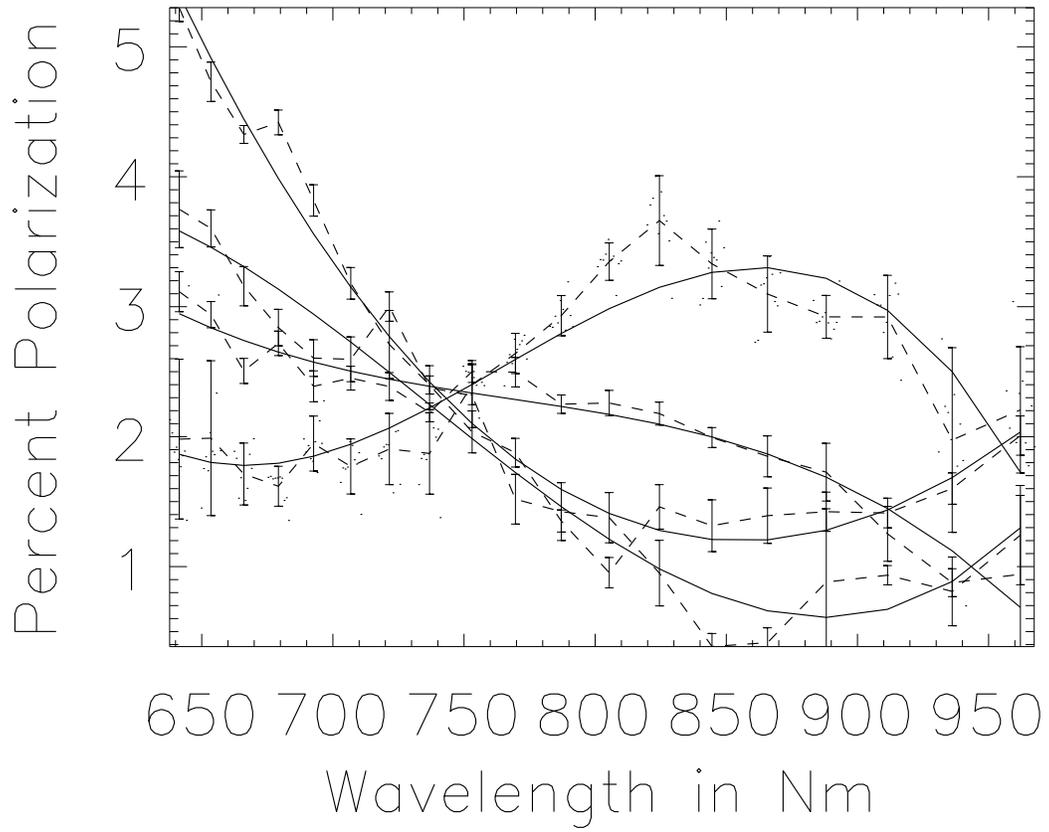}
\caption[pc]{\label{fig:142pc}
The degree of polarization in percent for the unpolarized standard star HD142373.  The progression in time at 650nm from the bottom curve to the top curve is 5:45, 7:20, 8:30, 7:50UT. }
\end{figure}

	The telescope-induced polarization for the first three of the standards in table \ref{upobs} are given in figures \ref{fig:125pc} through \ref{fig:142pc}.  Each star shows different polarization and illustrates how the induced polarization varies with wavelength and pointing since each star is unpolarized. These stars cover declinations of -7, +27, and +42, allowing us to measure the induced polarization at many different azimuths and altitude, from far north to far south.  
	
	The spectropolarimetry for HD125184 is shown in figure \ref{fig:125pc}.  It shows moderate change with wavelength ($\sim$1.5\% increase in the blue and decrease in the red).  The change in altitude is about 15$^\circ$ (63-49) and the azimuth changes by 55$^\circ$ (175-230).     

	HD114710 spectropolarimetry for many different pointings is shown in figure \ref{fig:114pc}.  The spectropolarimetry does not show much change with pointing, but it does show a strong change with wavelength at all pointings, varying from 1\% to 6\% in a roughly quadratic way.  The change in azimuth was 20$^\circ$ (76-44) and the altitude changed by 30$^\circ$ (309-290).

	Spectropolarimetry of HD142373, shown in figure \ref{fig:142pc} shows one of the strongest changes of polarization with pointing seen, but was in the north.  During the observations, the star's elevation does not change much ($\sim8^\circ$, 57-64) , but the stars azimuth changes by almost 100$^\circ$ (40-335) and the star transits in the middle of the data set, between 7:20 and 7:50UT.  The blue polarization (650nm) drops by 3\% while the 850nm polarization goes from near zero to 3\% and the 950nm rises by only 1\%.  

	Using these types of measurements for unpolarized standard stars at many different pointings, we can construct a response map for the telescope.

\subsection{The Telescope-Induced Polarization - An All-Sky Map}

	The observations of the degree of polarization for all the unpolarized standards observed, averaged to a single measurement per order are plotted on the alt-az plane to illustrate the induced polarization for a single order (wavelength) as a function of pointing.   Figure \ref{fig:unsurf0} shows the measurements for a single order (640nm).  There are 19 of these sky-map surfaces (19 orders in alt-az) that constitute the telescope response used for calibrating the instrument.  Note how the induced polarization reaches a maximum of $\sim$6\% at high altitude in the North and has a double-valley structure falling to around 2\% in the south (180$^\circ$) at lower elevations ($30^\circ$).  This order has the highest induced polarization meaning that the induced polarization is strongest at shorter wavelengths for this instrument. The longest wavelength order is shown in figure \ref{fig:unsurf18}.  The peak still occurs at high elevation in the north, but has a much lower magnitude of 3.5\% and falls to 1\% at low elevations in the south.  These maps illustrate the wavelength dependence of the polarization as a function of position on the sky.  As seen in figures 13-15, the errors are 0.1-1.0\% with a dependence on wavelength and pointing, making the calibration uncertain at this level.  
	To calibrate future spectropolarimetry we will use an interpolation of all 19 surfaces (wavelengths) to the pointing of the telescope in the middle of each data set as the induced polarization calibration applied to the measurements.  The error in this calculation is then calculated as the combined error in the nearest observations used in the interpolation.  We will do more observations to improve the calibration map.  It should be noted that for high-resolution line polarimetry, where one is only interested in the polarization changes across a spectral line (e.g. H$\alpha$), this calibration serves as the baseline.  Since the mirror-induced polarization effects do not change significantly over a single spectral line, the calibration process is simplified by removing the wavelength dependence of all the corrections.

\subsection{Zemax Model of Telescope Polarization - Mueller Matricies}

  We constructed a model of the telescope's polarization response using Zemax to compare the predicted effects with our observations.  Zemax software allows one to propagate arbitrarily polarized light through any optical design.  We wrote programs to compute the Mueller matrix of the telescope for any altitude or azimuth by tracking polarized light through our design of the telescope.  Using the optical constants for aluminum from the Handbook of Optical Constants of Solids II \citep{pal91}, we traced the polarization of the telescope to the first Coud\'e  focus, after the 2nd fold mirror in the optics room.  This allows us to get a qualitative idea of the polarization effects induced by the relative rotation of the altitude and azimuth mirror pairs ( m3-m4 and m6-m7 in figure \ref{fig:aeostel}).  Since the effect of aluminum oxide on the mirror surfaces is very significant and is variable in time, and the model does not include many of the common fore optics, this model is only an illustrative guide of what polarization the relative rotation of the mirrors will induce \citep{col92}.     

	To compute the Mueller matrix, we input pure polarization states $\pm$Q, $\pm$U, and $\pm$V at 400 pupil points, propagate the light through the telescope, and average the resulting polarized light at the image plane.  This is computed in 5$^\circ$ steps in altitude and azimuth giving 1296 independent measurements to project on to a sky-map.  The telescope induced polarization in unpolarized incident light is then just the IQ and IU Mueller matrix terms added in quadrature.  The result for 600nm light is shown in figure \ref{fig:ip}.  The induced polarization is expected to be minimal when the mirrors are crossed and maximal when they are aligned.  Since the azimuth axis has two angles where the mirrors are aligned (parallel and antiparallel),  we expect a double-ridge structure in azimuth.  The altitude axis mirrors are crossed at 0$^\circ$ so that the induced polarization rises with increasing altitude.  
	
\begin{figure}[!h,!t,!b]
\includegraphics[ width=.8\linewidth, angle=90]{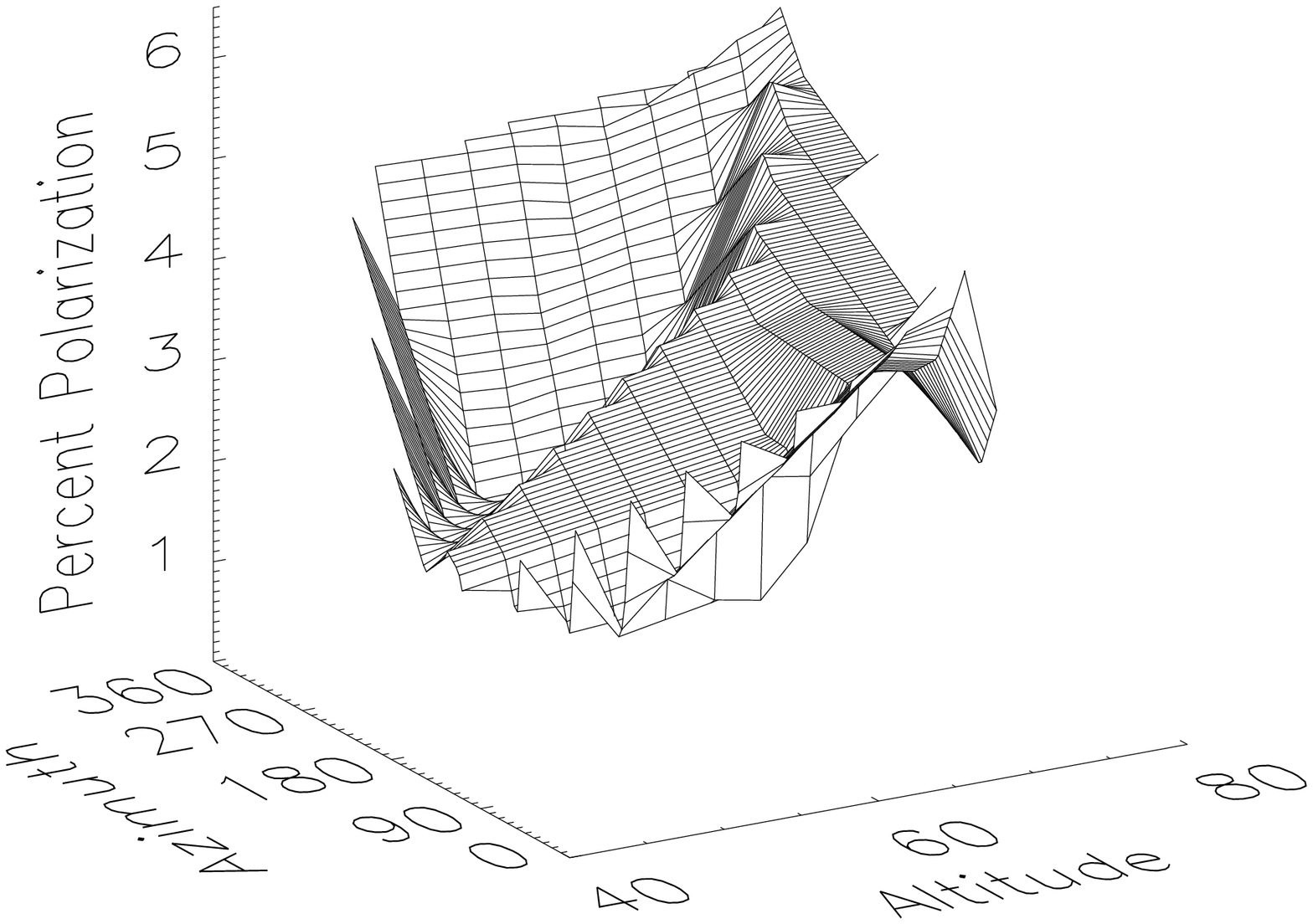}
\caption[unsurf0]{\label{fig:unsurf0}
The measured polarization of unpolarized standard stars for order 0 (640nm) in the alt-az plane. The peak is at 6\%.  The left axis is azimuth and the horizontal axis is altitude.  This is a triangulation of 19 independent observations.  The wavelength dependence for a few of the points in this plot are shown in figures \ref{fig:125pc} through \ref{fig:142pc}.  }   
\end{figure}

\begin{figure}[!h,!t,!b]
\includegraphics[ width=.8\linewidth, angle=90]{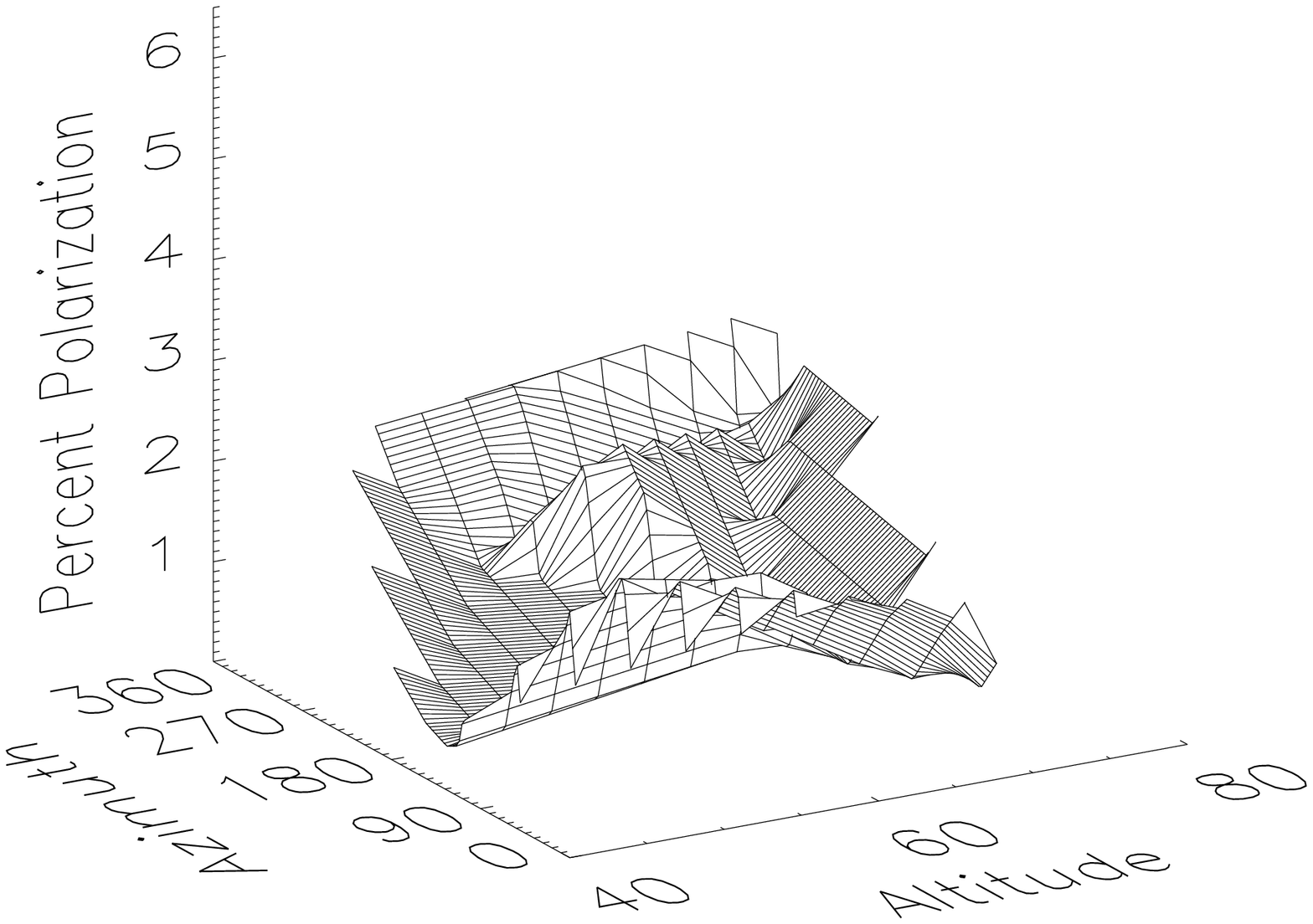}
\caption[unsurf18]{\label{fig:unsurf18}
The measured polarization of unpolarized standard stars for order 18 (960nm) in the alt-az plane. The peak is near 4\%.  The left axis is azimuth and the horizontal axis is altitude.  This is a triangulation of 19 independent observations.  The wavelength dependence for a few of the points in this plot are shown in figures \ref{fig:125pc} through \ref{fig:142pc}.  }   
\end{figure}

\begin{figure}[!h,!t,!b]
\includegraphics[width=.8\linewidth, angle=90]{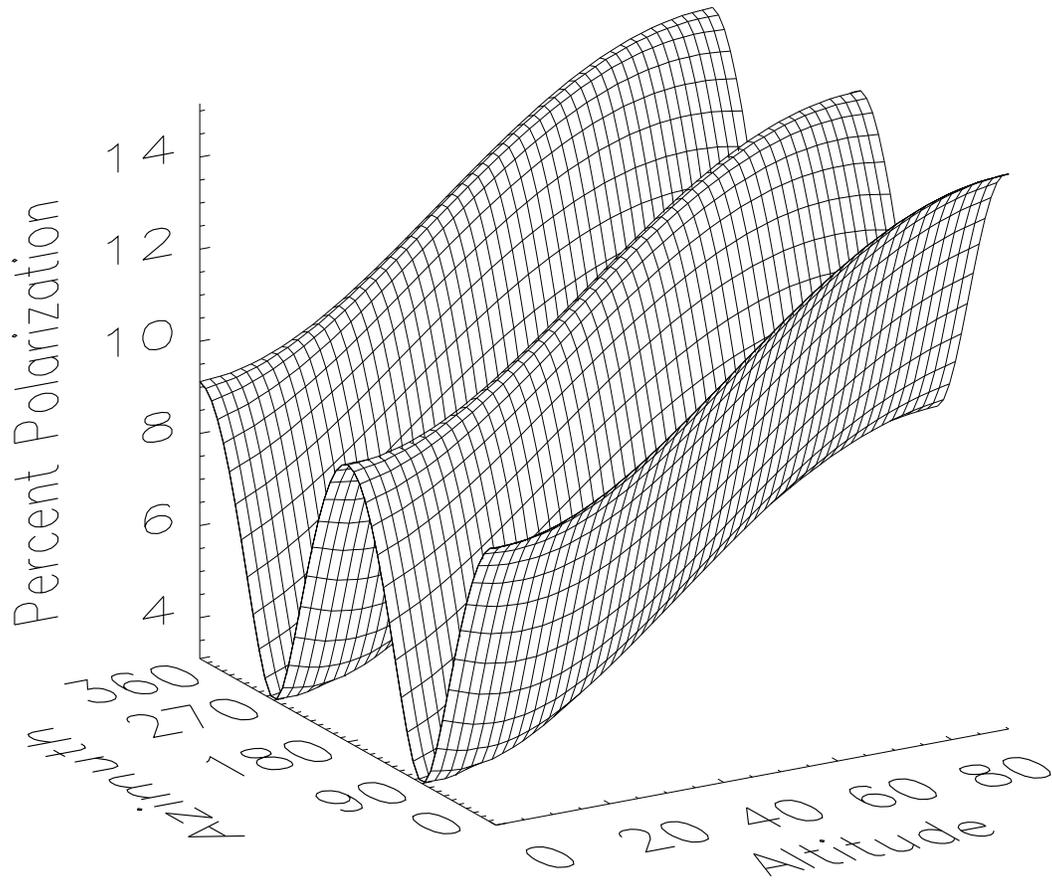}
\caption[iq]{\label{fig:ip}
Zemax model of the telescope induced polarization using bare aluminum surfaces - the predicted induced polarization projected on the sky at 600nm. }   
\end{figure}

	This model compares favorably with the measurements in figure \ref{fig:unsurf0}.  Since we have only obtained 19 independent measurements of unpolarized standard stars, the sparsely sampled map in figure \ref{fig:unsurf0} is not expected to match this much higher resolution model.  The measured polarization shows a roughly double ridge structure in azimuth which rises with altitude, as expected.  The measured polarization is roughly half the model's prediction, but this is entirely plausable given that the effect of the oxide layer on the mirrors can be drastic.  We have computed this model for many wavelengths, and all show this qualitative shape.
	
	The model also gives justification for neglecting circular polarization.  The circular polarization cross-talk terms, Mueller matrix elements IV and VI, are less than 5\%.  The VQ and VU terms, which describe incident circular polarization becoming measured linear polarization are larger, typically 0-70\%, as are the QV and UV terms.  Since most astronomical objects have circular polarization that is an order of magnitude smaller than linear polarization, we do not anticipate significant calibration problems with this amount of circular to linear cross-talk.

	The last of the model predictions is that the QI and UI terms are small (0 to 20\%), leading to low depolarization, and the rotation of the plane of polarization does not vary much with wavelength (QU and UQ terms change by a few \%).  
	
	Since the actual polarization properties of the mirrors are very sensitive to the optical constants of the aluminum, and the aluminum oxide coating on all mirrors is not included in our models, these predictions are not expected to be quantitatively precise.  They do however, give a rough guide to the altitude, azimuth, and wavelength dependence.  We have yet to monitor the polarization of the telescope as a function of time, to test the stability of these calibrations, and hope to do this in the coming months.

\subsection{Scattered Sunlight - A Linearly Polarized Source for Calibration}

	At twilight, scattered sunlight is a linearly polarized source with a roughly known position angle and degree of polarization at all pointings.   This assumes singly-scattering of incident solar radiation.  Singly-scattered light is polarized orthogonal to the scattering plane with the degree of polarization proportional to $sin^2 \theta$  \citep{cou88}.  Since the scattering geometry changes with pointing, so does the degree of polarization of the scattered sunlight (which reaches a maximum 90$^\circ$ from the sun).  Observing scattered sunlight at many different altitudes and azimuths allows us to check the instrument's response to linearly polarized light.  This scattered light is not a perfect source however.  The degree of polarization is also a strong function of the scattering properties of the air.  In particular, aerosol content (salt, dust, molecules, water, etc), optical depth (altitude and horizon distance) and reflections from the Earth's surface (land and ocean) can significantly change the observed linear polarization \citep{lee98, liu97, suh04, cou88, cro05}.  Our observations were done at high altitude, surrounded by a reflecting ocean surface.  Since the sun was up during some of our observations, the polarized light reflected off the ocean surface complicates our measurements.  The broad trends in the degree of polarization we measured are in agreement with the singly-scattered sky polarization model, but the linear polarization observed by many researchers has been shown to be functions of time, atmospheric opacity, aerosol size distributions, and composition, all of which vary significantly from day to day \citep{cou88, liu97, suh04}.  Since we have no accurate, quantitative model of the atmospheric polarization to compare with our observations, we will use them only as an illustration of the types of effects one can expect from a highly polarized source. 

\begin{figure}[!h,!t,!b]
\includegraphics[ width=.8\linewidth, angle=90]{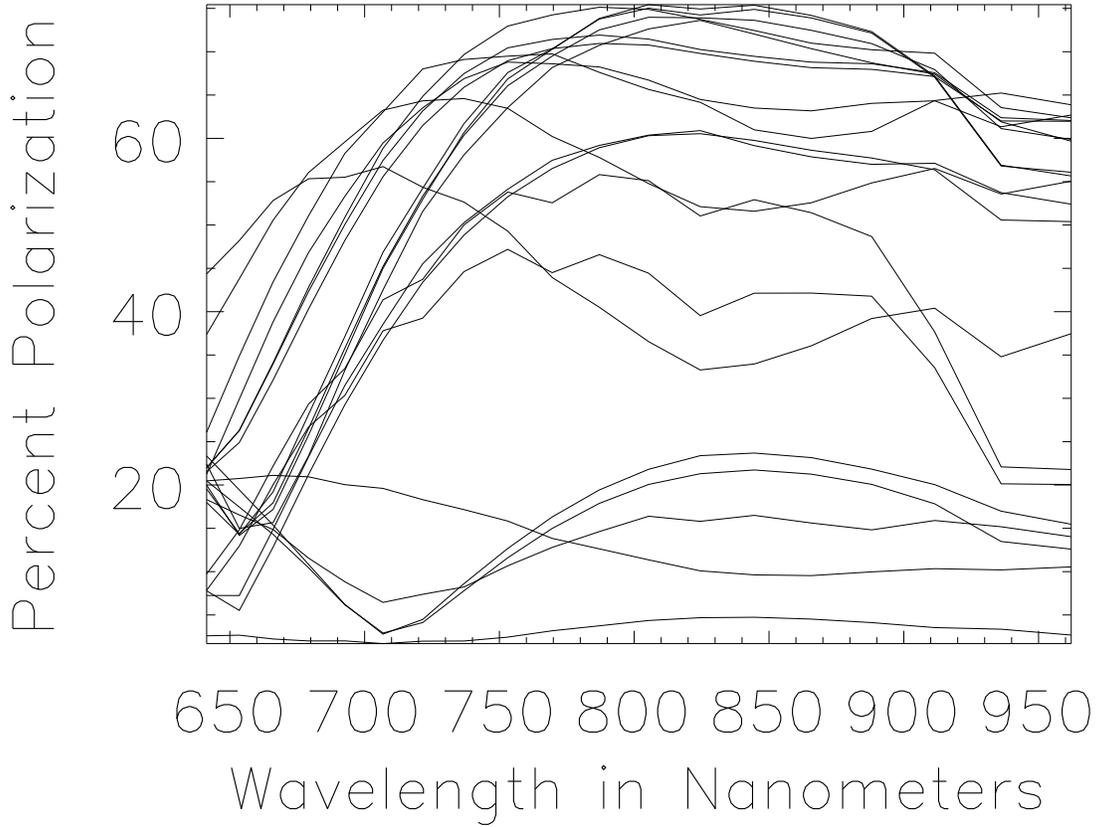}
\caption[skyp]{\label{fig:skyp}
The fractional degree of polarization of scattered sunlight at combinations of altitudes [30,50,65,75,90] and azimuths [N,E,S,W].  The degree of polarization of the scattered sunlight is intrinsically a function of wavelength and pointing since the scattering geometry changes, making quantatative interpretation of the HiVIS degree of polarization difficult, but there is good correlation with theoretical skylight polarization models.}
\end{figure}

\begin{figure}[!h,!t,!b]
\includegraphics[ width=.8\linewidth, angle=90]{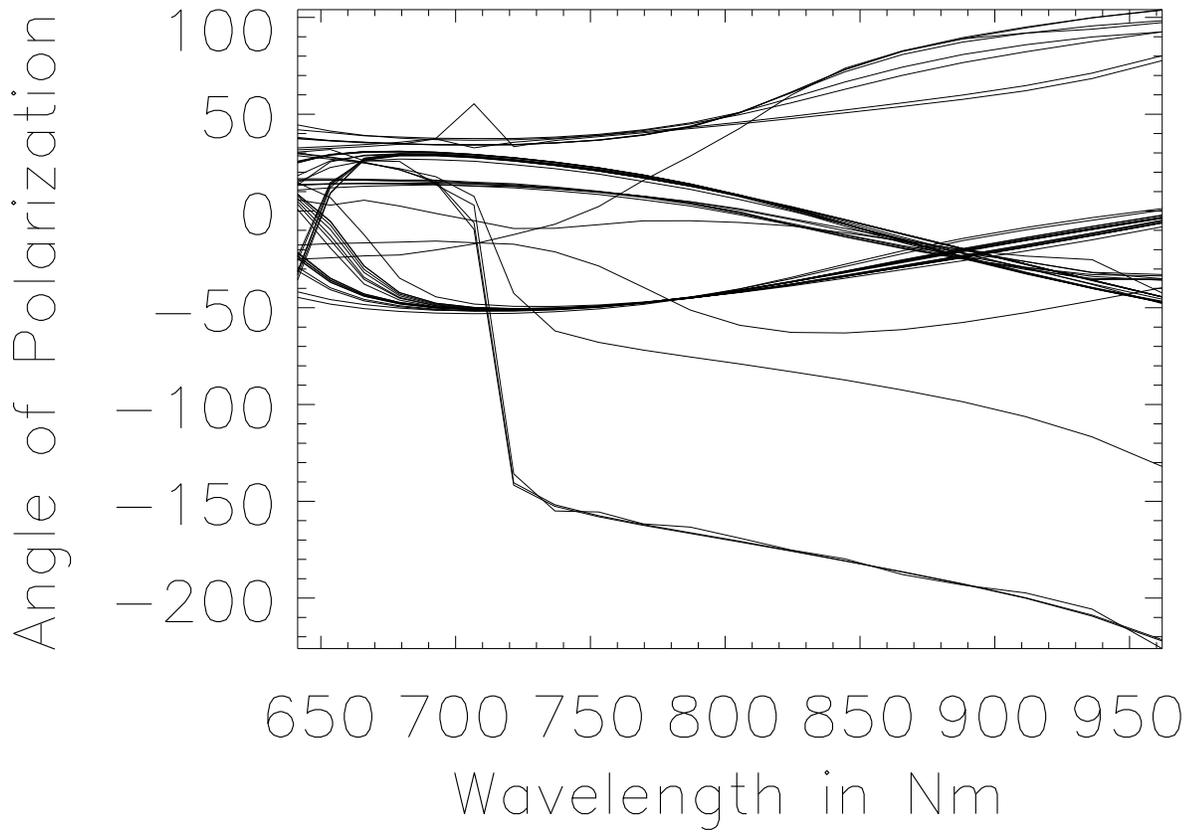}
\caption[sky_theta]{\label{fig:skytheta}
The calculated angle of polarization ($\frac{1}{2} tan^{-1}\frac{Q}{U}$) for scattered sunlight at combinations of altitudes [30,50,65,75,90] and azimuths [N,E,S,W].}
\end{figure}

	We obtained 52 polarization measurements (416 spectra) of scattered sunlight at altitudes and azimuths of [30$^\circ$, 50$^\circ$, 65$^\circ$, 75$^\circ$, 90$^\circ$] and [0$^\circ$, 90$^\circ$, 180$^\circ$, 270$^\circ$] to create a map of the degree of polarization and position angle for the sky to compare with other observations  \citep{hor98, cou88, hor02}.  These spectra show varied wavelength dependences.  A plot of the degree of polarization for some of the 52 sky polarization measurements is shown in figure \ref{fig:skyp}, and the position angle is shown in figure \ref{fig:skytheta}.  The rotation of the plane of polarization was a very strong function of wavelength and pointing, with typically 90$^\circ$ of rotation from order 0 through 18.   

\begin{figure}[!h,!t,!b]
\includegraphics[ width=.8\linewidth, angle=90]{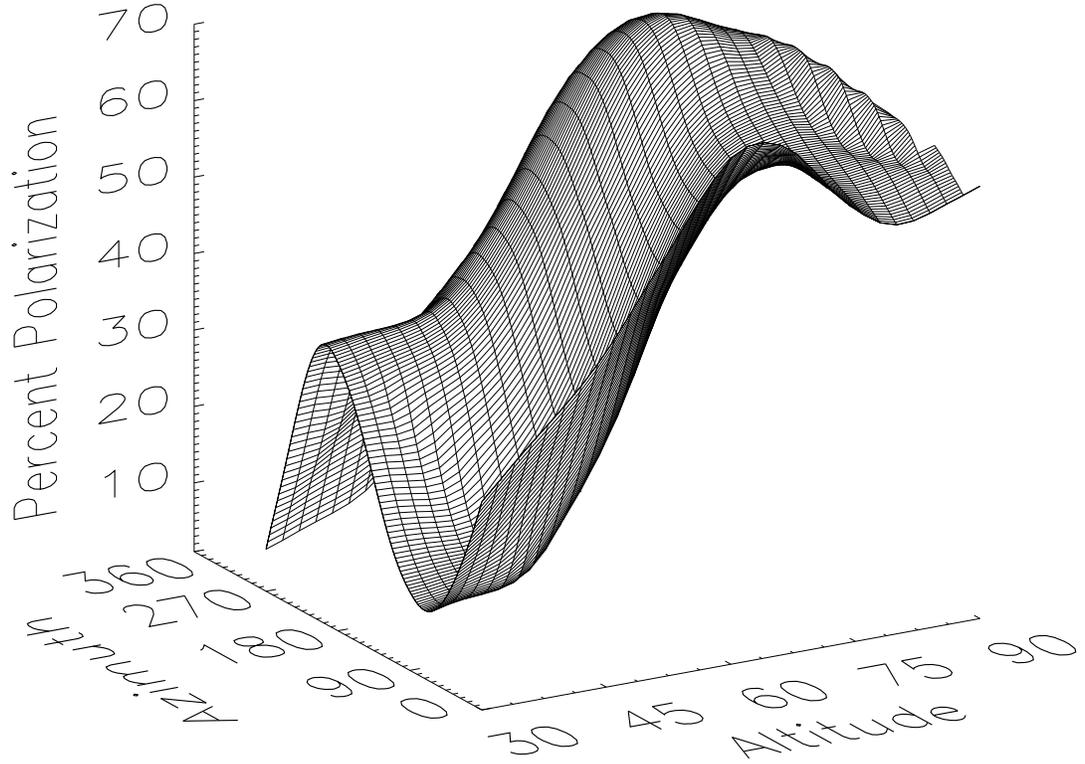}
\caption[skysurf]{\label{fig:skysurf}
The measured polarization of scattered sunlight at sunset for order 14 (860-870nm) in the alt-az plane.  At sunset polarization is maximum 90$^\circ$ off the sun (N, S, Zenith) and minimum at the anti-solar point (low altitude in the East and West). }   
\end{figure}

	As with the unpolarized standards, the scattered sunlight polarization measurements can be plotted on the altitude-azimuth plane.  Figure \ref{fig:skysurf} shows the measured degree of polarization projected on the sky for order 14 (860-870nm).  Since scattered sunlight reaches a maximum polarization 90$^\circ$ away from the sun, and these measurements were done at sunset, the degree of polarization is maximum along azimuths of 0$^\circ$ and 180$^\circ$ and minimum at 90$^\circ$ and 270$^\circ$.  The figure shows a very pronounced structure that is consistent with a simple single-scattering (Rayleigh) atmosphere \citep{cou88}.  The polarization is highest at high altitudes and in the north and south.  However, the measured polarization seen in figure \ref{fig:skyp} does show strong wavelength dependence and so these surfaces change significantly with wavelength.  This is entirely plausible since the scattering properties of the atmosphere are also strong functions of wavelength.  

 	The strong rotation of the plane of polarization in figure \ref{fig:skytheta} is also very interesting.  Since the reflections off the ocean surface are very wavelength dependent, we cannot say how much of this rotation is instrumental and how much is ocean reflection.   This rotation is significantly different from that observed in other locations, but our environment is also very different \citep{hor98, cou88, hor02}.  The rotation is strongest when the degree of polarization is lowest, suggesting a mixture of the singly-scattered light, with a smaller amount reflected light (compare figures \ref{fig:skyp} and \ref{fig:skytheta}).

\subsection{Calibration Summary}

	We have illustrated the polarization effects of a altitude azimuth telescope using the Coud\`e focus.  The moving mirrors can induce, reduce, and rotate polarization as functions of position, pointing, and wavelength.  To do a complete calibration of the induced polarization, measurements of unpolarized sources must be made at many pointings so that the source of interest can be calibrated for the telescope polarization.  Our measurements with a linear polarizer mounted behind the spectropolarimeter showed that we can detect linear polarization downstream of the spectropolarimeter with an efficiency better than 95\% in the useful range of the linear polarizer.  The unpolarized standard star measurements showed that the telescope induces polarization of order 1-6\% with a systematic variation in pointing and wavelength.  The induced polarization is higher at high elevations, has a double-ridge structure in azimuth, and falls with wavelength, consistent with the Zemax polarization model of the telescope we constructed. The unpolarized star measurements have been projected back onto the sky so that a map of the telescope response has been made for the induced polarization to allow us to calibrate other sources of interest.  The scattered sunlight degree of polarization measurements are consistent in broad form with the single-scattering Rayleigh atmosphere, giving us confidence that we can detect linear polarization across all wavelengths.  The plane of polarization measurements were complicated by our site location near a strongly reflecting surface (ocean).  The rotation was a strong function of wavelength and pointing, likely due to mixing of different polarized sources, and we were unable to measure the telescopes rotation of the plane of polarization.  The Zemax model predicts the rotation to be a small function of wavelength  (a few \% in the QQ, QU, UQ, UU Mueller matrix terms).

\section{Conclusions}

	This was the first test of the HiVIS spectropolarimeter over broad wavelengths.  We have illustrated a method for calibrating an alt-az coud\'e  telescope with many moving mirrors for induced polarization, efficiency of polarization detection, and rotation of the plane of polarization.  Measurements of unpolarized sources and linearly polarized sources ($\sim1000$ spectra) with known position angles at many pointings were necessary to obtain good altitude-azimuth coverage.  The AEOS telescope shows a maximum of 6\% induced polarization at 640nm and high altitude with values falling with wavelength and altitude to near 1\%.  We constructed an all-sky map of the telescope-induced polarization and can use it to calibrate future observations.  We hope to do more observations of unpolarized and polarized sources to improve the resolution of the calibration sky maps.

\clearpage

\begin{table}[!h,!t,!b]
\normalsize
 \begin{center}
\caption{{Unpolarized Standard Star Observation}\label{upobs}}
\begin{tabular}{lcccccrr}
\tableline\tableline
Star & RA (hm) & Dec (dm) &  Date(UT) & Time(UT) & Exp(s) & Alt & Azi \\
\tableline
\tableline
HD114710 & 13 12 & +27 53 & 702 & 5:40 & 10 & 76 & 309  \\
HD114710 & 12 12 & +27 53 & 701 & 5:40 & 10 & 76 & 309  \\
HD114710 & 13 12 & +27 53 & 630 & 5:55 & 20 & 73 & 302  \\
HD114710 & 13 12 & +27 53 & 630 & 7:00 & 20 & 60 & 292  \\
HD114710 & 13 12 & +27 53 & 702 & 7:15 & 10 & 56 & 291  \\
HD114710 & 13 12 & +27 53 & 701 & 8:10 & 10 & 44 & 290  \\
\tableline
HD125184 & 14 15 & -07 19 & 701 & 5:45 & 30 & 63 & 175  \\
HD125184 & 14 15 & -07 19 & 703 & 5:55 & 30 & 61 & 180  \\
HD125184 & 14 15 & -07 19 & 702 & 7:20 & 30 & 56 & 218  \\
HD125184 & 14 15 & -07 19 & 701 & 8:00 & 30 & 49 & 230  \\ 
\tableline
HD142373 & 15 51 & +42 35 & 702 & 5:45 & 20 & 57 & 040  \\
HD142373 & 15 51 & +42 35 & 630 & 7:20 & 10 & 67 & 006  \\
HD142373 & 15 51 & +42 35 & 701 & 7:50 & 10 & 67 & 350  \\
HD142373 & 15 51 & +42 35 & 701 & 8:30 & 10 & 64 & 335  \\
\tableline
HD154345 & 17 01 & +47 08 & 702 & 5:50 & 30 & 45 & 043  \\
HD154345 & 17 01 & +47 08 & 701 & 8:40 & 30 & 63 & 002  \\
HD154345 & 17 01 & +47 08 & 704 & 8:50 & 15 & 63 & 035  \\
\tableline 
HD165908 & 18 05 & +30 33 & 703 & 5:40 & 30 & 57 & 065  \\
HD165908 & 18 05 & +30 33 & 702 & 7:30 & 30 & 34 & 066  \\
\tableline
\tableline
\end{tabular}
\end{center}
\end{table}

\clearpage

\clearpage

\clearpage

\clearpage

\clearpage

\clearpage

\clearpage

\clearpage

\clearpage

\clearpage

\clearpage

\clearpage

\clearpage

\clearpage

\clearpage

\clearpage

\clearpage

\clearpage

\clearpage

\clearpage

\end{document}